\documentclass[twocolumn, 10pt]{article}
\usepackage{authblk, units}
\usepackage[margin=1.5cm]{geometry}

\usepackage{graphicx}
\usepackage{sidecap}
\usepackage{amssymb,amsmath, bbm}
\usepackage{epstopdf}
\usepackage{color}
\usepackage[usenames,dvipsnames]{xcolor}
\usepackage{mathtools}
\usepackage{floatrow}
\newfloatcommand{capbtabbox}{table}[][\FBwidth]

\newcommand{\bra}[1]{\ensuremath{\left\langle #1\right|}}   
\newcommand{\ket}[1]{\ensuremath{\left|#1\right\rangle}}   
\newcommand{\mean}[1]{\ensuremath{\langle{#1}\rangle}}
\usepackage{braket}
\newcommand{\B}[0]{\ensuremath{\mathrm{B}}}
\newcommand{\bs}[1]{\vspace{0.2cm}\noindent\textbf{#1}}
\newcommand{\bss}[1]{\noindent\textbf{#1}}

\begin{document}

\title{Violating Bell's inequality with an artificial atom and
a cat state in a cavity}
\author[$*$,1]{Brian\ Vlastakis}
\author[$*$,1]{Andrei\ Petrenko}
\author[1]{Nissim\ Ofek}
\author[1]{Luyan\ Sun}
\author[1]{Zaki\ Leghtas}
\author[1]{Katrina\ Sliwa}
\author[1]{Yehan\ Liu}
\author[1]{Michael\ Hatridge}
\author[1]{Jacob\ Blumoff}
\author[1]{Luigi\ Frunzio}
\author[1,2]{Mazyar\ Mirrahimi}
\author[1]{Liang\ Jiang}
\author[1]{M.\ H.\ Devoret}
\author[1]{R.\ J.\ Schoelkopf}
\affil[1]{Departments of Physics and Applied Physics, Yale University, New Haven, CT 06510, USA}
\affil[2]{INRIA Paris-Rocquencourt, Domaine de Voluceau, B.P. 105, 78153 Le Chesnay Cedex, France}
\affil[$*$]{These authors contributed equally to this work.}
\twocolumn[
  \begin{@twocolumnfalse}
    \maketitle
\begin{abstract}
The `Schr\"{o}dinger's cat' thought experiment highlights the counterintuitive facet of quantum theory that entanglement can exist between microscopic and macroscopic systems, producing a superposition of distinguishable states like the fictitious cat that is both alive and dead. The hallmark of entanglement is the detection of strong correlations between systems, for example by the violation of Bell's inequality~\cite{NielsenQI}. Using the CHSH variant~\cite{Clauser:1969hk} of the Bell test, this violation has been observed with photons~\cite{Freedman:1972vr,Aspect:1981wl}, atoms~\cite{Rowe:2001ic,Hofmann:2012jb}, solid state spins~\cite{Pfaff:2012gy}, and artificial atoms in superconducting circuits~\cite{Ansmann:2009eq}. For larger, more distinguishable states, the conflict between quantum predictions and our classical expectations is typically resolved due to the rapid onset of decoherence. To investigate this reconciliation, one can employ a superposition of coherent states in an oscillator, known as a cat state~\cite{book:haroche06}. In contrast to discrete systems, one can continuously vary the size of the prepared cat state and therefore its dependence on decoherence. No violation of Bell's inequality has yet been observed for a system entangled with a cat state. Here we demonstrate and quantify entanglement between an artificial atom and a cat state in a cavity, which we call a `Bell-cat' state. We use a circuit QED~\cite{Wallraff:2004dy} architecture, high-fidelity measurements, and real-time feedback control to violate Bell's inequality~\cite{Clauser:1969hk} without post-selection or corrections for measurement inefficiencies. Furthermore, we investigate the influence of decoherence by continuously varying the size of created Bell-cat states and characterize the entangled system by joint Wigner tomography. These techniques provide a toolset for quantum information processing  with entangled qubits and resonators~\cite{Mirrahimi:2014js}. While recent results have demonstrated a high level of control of such systems~\cite{Vlastakis:2013ju,Sun:2014ha,Hofheinz:2009ba}, this experiment demonstrates that information can be extracted efficiently and with high fidelity, a crucial requirement for quantum computing with resonators~\cite{DiVincenzo:2000}.
\end{abstract}
  \end{@twocolumnfalse}
]
\begin{figure*}[t]
\centering
\includegraphics[]{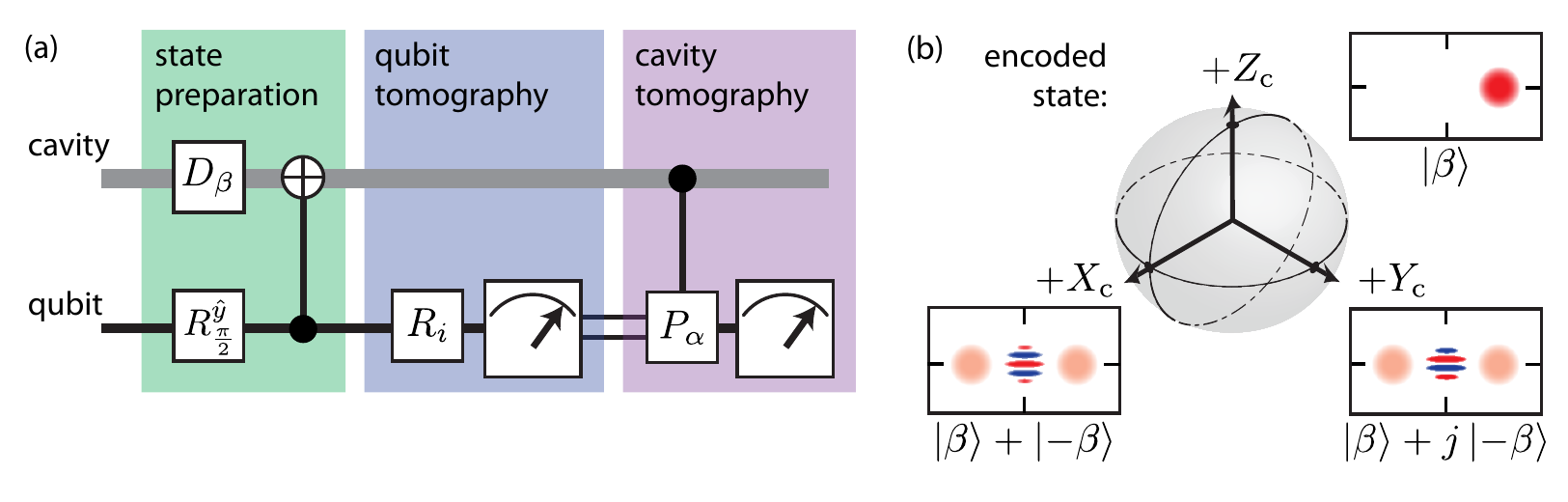}
\caption{\textbf{Sequential detection for entanglement characterization.}
    (a)~A quantum circuit outlines the method to prepare and measure entanglement between a qubit and cavity state using sequential detection. State preparation is performed by first creating a product state $\ket{\psi} = \tfrac{1}{\sqrt{2}}(\ket{g}+\ket{e})\otimes\ket{\beta}$ with a cavity displacement $D_\beta$ of amplitude $\beta$ and a qubit gate $R^{\hat y}_{\frac{\pi}{2}}$ corresponding to a $\tfrac{\pi}{2}$ rotation around the $\hat y$-axis. A conditional gate using the dispersive interaction, produces the entangled state $\ket{\psi_\mathrm{B}} = \tfrac{1}{\sqrt{2}}(\ket{g,\beta}+\ket{e,{-\beta}})$. Tomography is performed by measuring an observable of both the qubit and cavity with sequential QND measurements. A pre-rotation $R_i$ allows qubit detection along one of three basis vectors $X$, $Y$, and $Z$. The qubit is reset and a cavity observable, the displaced photon number parity $P_\alpha$, is mapped to the qubit for a subsequent measurement. Sequential detections are binary results compared shot-by-shot to determine qubit-cavity correlations. (b)~The space spanned by the superposition of quasi-orthogonal coherent states $\ket{\beta},~\ket{{-\beta}}$ constitutes an encoded quantum bit in the cavity. While the cavity state can be represented by its Wigner function, this logical state is also described by a vector within its encoded Bloch sphere. For well-separated coherent state superpositions, the entangled state $\ket{\psi_\mathrm{B}}$ is then equivalent to a two-qubit Bell state.
}
   \label{fig:1}
\end{figure*}

Quantum information processing necessitates the creation and detection of complex entangled states. Many implementations aim to encode quantum information into a register of physical qubits.  Alternative encoding schemes using cat states take advantage of a cavity resonator's large Hilbert space, and allow redundant qubit encodings that simplify the operations needed to initialize, manipulate, and measure the encoded information~\cite{Gottesman:2001jb,Leghtas:2013wx}. The cavity mode's state can be completely described by direct measurements in the continuous-variable basis with the cavity state Wigner function~\cite{Lutterbach:1997cn}. We extend this concept to express an entangled qubit-cavity state in what we call the joint Wigner representation. We construct this representation by performing a sequence of two subsequent quantum non-demolition (QND) measurements~(Fig.~\ref{fig:1}), where a qubit state measurement is correlated with a subsequent cavity state measurement. When working in a cavity subspace, however, complete state tomography may not be required and in fact many fewer measurements could be used to determine a state, such as one with a clear mapping to single-qubit observables. By choosing an encoding scheme where logical states of a quantum bit are mapped onto a superposition of coherent states $\ket{\beta}$ and $\ket{{-\beta}}$, we can condense the joint Wigner representation down to just sixteen correlations, equivalent to a two-qubit measurement set. Using direct fidelity estimation~\cite{daSilva:2011ej,Flammia:2011dfe} and CHSH Bell test witnesses~\cite{Park:2012iw} within this logical basis, we assess the degree of entanglement of the state. We investigate this system's susceptibility to decoherence by continuously increasing the cat state amplitude $|\beta|$. We measure a range in which correlations surpass the Bell violation threshold and observe its decline due to decoherence, benchmarking the efficiency of our encoding and detection schemes with cat-state qubits.

This experiment utilizes a circuit QED architecture~\cite{Wallraff:2004dy,Paik:2011hd} consisting of two waveguide cavities coupled to a single transmon qubit~\cite{Kirchmair:el5H1JZQ,Vlastakis:2013ju,Sun:2014ha}. One long-lived cavity (relaxation time $\tau_s = 55~\mathrm{\mu s}$) is used for quantum information storage, while the other cavity, with fast field decay (relaxation time $\tau_r = 30~\mathrm{ns}$ ), is used to realize repeated measurements. A transmon qubit (relaxation and decoherence times $T_1,~T_2 \approx 10~\mathrm{\mu s}$) is coupled to both cavity modes and mediates entanglement and measurement of the storage cavity state. All modes have transition frequencies between $5\text{--}8~\mathrm{GHz}$ and are off-resonantly coupled (see methods for details). The storage cavity and qubit mode are well described by the dispersive Hamiltonian:
\begin{equation}
    \label{eq:hamiltonian}
H/\hbar = \omega_s a^\dagger a + (\omega_q  - \chi a^\dagger a ) \ket{e}\bra{e}
\end{equation}
where $a$ is the storage cavity ladder operator, $\ket{e}\bra{e}$ is the excited state qubit projector, $\omega_s,~\omega_q$ are the storage cavity and qubit transition  frequencies, and $\chi$ is the dispersive interaction strength between the two modes ($1.4~\mathrm{MHz}$). This interaction creates a shift in the transition frequency of one mode dependent on the other's excitation number, resulting in qubit-cavity entanglement~\cite{Brune:1992cp}. As described in Fig.~\ref{fig:1}, the system is first prepared in a product state $\ket{\psi}=\frac{1}{\sqrt{2}}(\ket{g} + \ket{e})\otimes \ket{\beta}$, where $\ket{g},~\ket{e}$ are the ground and excited states of the qubit and $\ket{\beta}$ is a coherent state of the cavity mode. Under the dispersive interaction we allow the system to evolve for a time $t = \frac{\pi}{\chi}$, creating the entangled state:
\begin{equation}
    \label{eq:bell_cat}
\ket{\psi_\mathrm{B}} = \tfrac{1}{\sqrt{2}}(\ket{g,\beta}+\ket{e,{-\beta}})
\end{equation}
which we call a Bell-cat state~\cite{Brune:1996bv,Vlastakis:2013ju,Sun:2014ha}, mirroring the form of a two-qubit Bell state (e.g. $\ket{\psi} = \tfrac{1}{\sqrt{2}} ( \ket{gg} + \ket{ee})$).  

\begin{figure*}[t]
\centering
\includegraphics[]{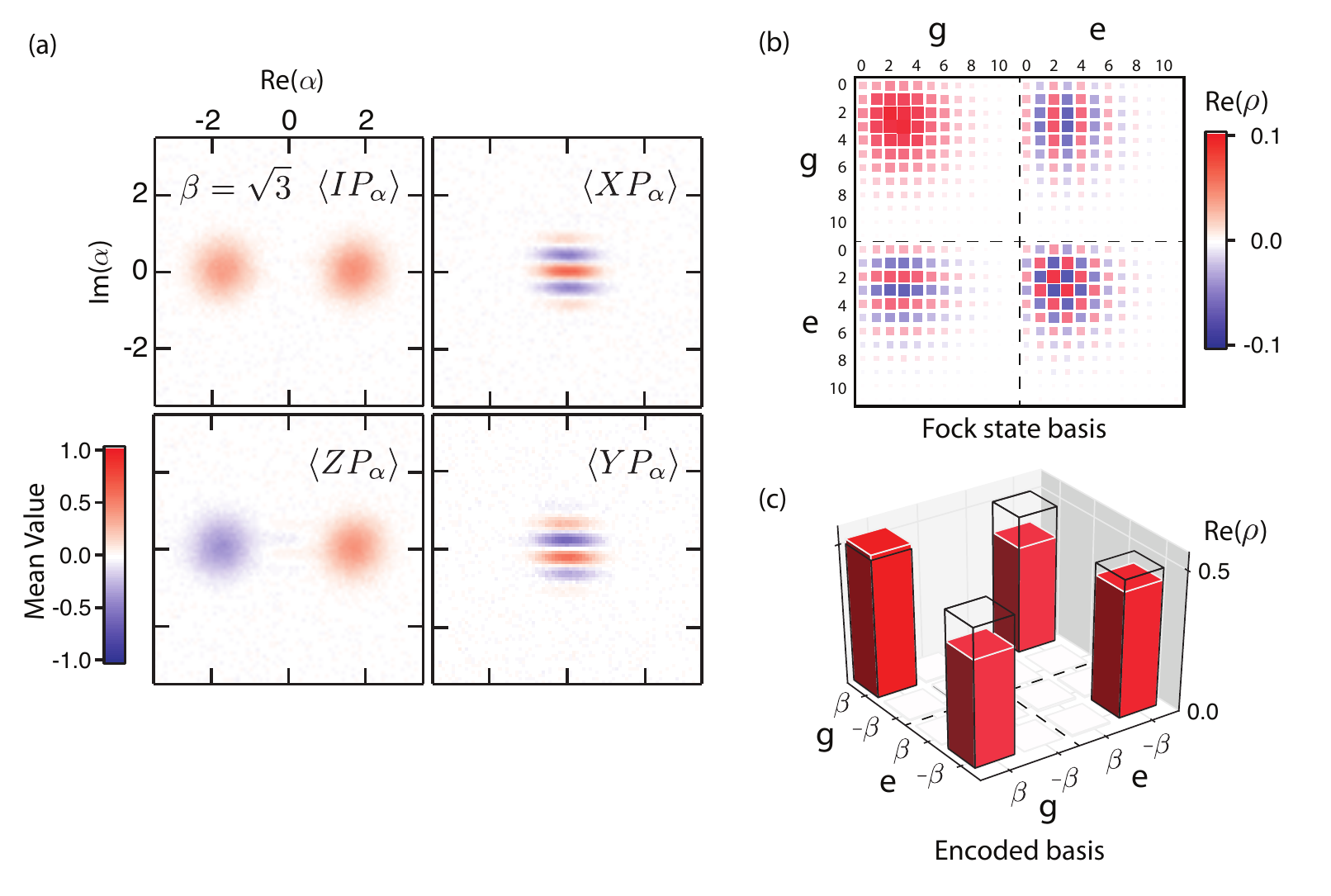}
\caption{\textbf{Joint Wigner tomography of a Bell-cat state. }
    (a) The set of joint Wigner functions $W_i(\alpha) = \tfrac{2}{\pi}\braket{\sigma_i P_\alpha}$ represents the state of a qubit-cavity system with correlations between the qubit $\sigma_i = \{I~,X~,Y~,Z\}$ and cavity $P_\alpha$ reported for a state $\ket{\psi_\mathrm{B}}$ and displacement amplitude $\beta = \sqrt{3}$. Shown are measurements comprised of four grids of 6500 correlations between the qubit and cavity states. Interference fringes in $\braket{XP_\alpha}$ and $\braket{YP_\alpha}$ reveal quantum coherence in the entangled state. (b) A density matrix reconstruction shows the combined qubit-cavity state $\rho$ in the Fock state basis. (c) Projecting onto the logical basis $\ket{\beta}\bra{\beta} + \ket{{-\beta}}\bra{{-\beta}}$, this state can be further reduced exhibiting the traditional Bell state form.
}
   \label{fig:2}
\end{figure*}
Correlating sequential high-fidelity measurements of the qubit and cavity allows state tomography of the composite system. We use a Josephson bifurcation amplifier~\cite{Vijay:2009ko} in a double-pumped configuration in combination with a dispersive readout to perform repeated QND measurements with qubit detection fidelity of 98.0\% at a minimum of 800~ns between measurements. The first measurement detects the qubit along one of its basis vectors $\{ X,~Y,~Z \}$. This value is recorded and the qubit is reset to $\ket{g}$ using real-time feedback (see methods). The displaced photon-number parity observable $P_\alpha$ of the cavity is subsequently mapped onto the qubit using Ramsey interferometry~\cite{Lutterbach:1997cn} before a second qubit state detection. The cavity observable $P_\alpha = D_\alpha P D_\alpha^\dagger$ where $D_\alpha$ is the displacement operator and $P$ the photon number parity operator is detected with 95.5\% fidelity (see methods). The Wigner function $W(\alpha) = \frac{2}{\pi}\braket{P_\alpha}$ is constructed from an ensemble of such measurements with different displacement amplitudes $\alpha$. The correlations between the qubit and cavity states make up what we refer to as the joint Wigner functions:
\begin{equation}
W_i(\alpha) = \tfrac{2}{\pi}\braket{\sigma_i P_\alpha}
\end{equation}
where $\sigma_i$ is an observable in the qubit Pauli set $\{I,~X,~Y,~Z\}$. These four distributions are a complete representation of the combined qubit-cavity quantum state (see Fig.~\ref{fig:2}). While other representations exist for similar systems~\cite{Eichler:2012vy,Morin:2014,Jeong:2014bl,LinPeng:2013eo}, $W_i(\alpha)$ is directly measured with this detection scheme and does not require a density matrix reconstruction. By an overlap integral (see methods), we determine the fidelity to a target state $\mathcal{F} = \bra{\psi_\B}\rho \ket{\psi_\B} = \frac{\pi}{2} \sum_i \int W^\B_i(\alpha) W_i(\alpha) \mathrm{d}^2 \alpha$ where $W^\B_i(\alpha)$ are the joint Wigner functions of the ideal state $\ket{\psi_\mathrm{B}}$ and $W_i(\alpha)$ are the measured joint Wigner functions (normalized), yielding a state fidelity $\mathcal{F} = (87 \pm 2)\% $ for a displacement amplitude $\beta = \sqrt{3}$. This amplitude was chosen to ensure orthogonality between logical states $|\langle \beta | {-\beta} \rangle |^2 = 6 \times 10^{-5} \ll 1$ with minimal trade-off due to photon loss. Furthermore, the efficiency of our detection scheme can be quantified by the visibility of the unnormalized joint Wigner measurements $\mathcal{V} = \frac{2}{\pi}\int \mean{IP_\alpha} \mathrm{d}^2 \alpha = (85 \pm 1)\%$. Visibility $\mathcal{V}$ is primarily limited by measurement fidelity and qubit decoherence between detection events (see methods). The parameters $\mathcal{F}$ and $\mathcal{V}$ represent critical benchmarks for creating and retrieving information from entangled states.

\begin{SCfigure*}
\centering
\includegraphics[scale = 0.8]{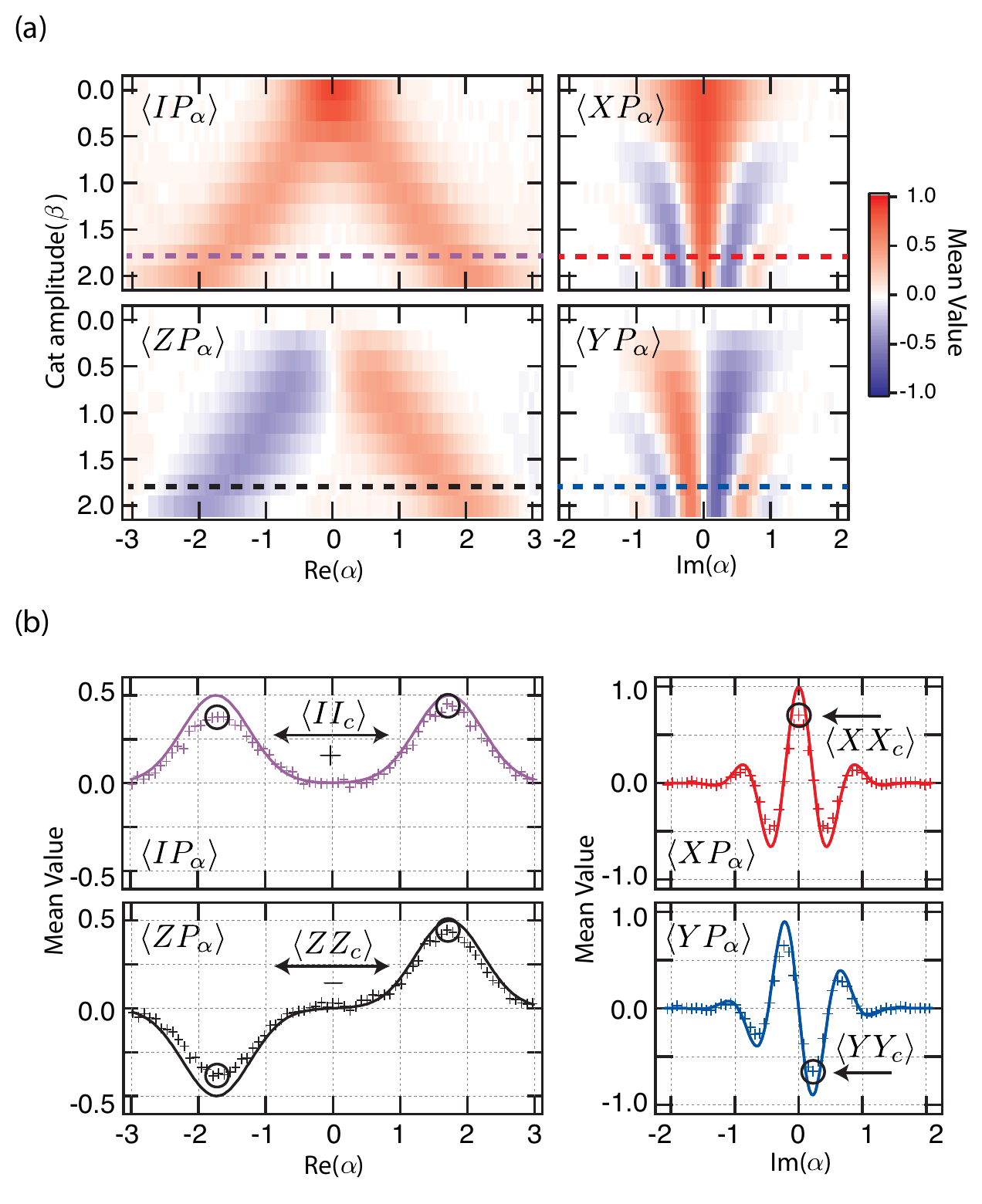}
\caption{\textbf{Qubit-cavity correlations.}
    (a) Correlations are measured for entangled states $\ket{\psi_\B}$ with displacement amplitudes ranging from $\beta=0$ to 2. Cuts in joint Wigner functions $\braket{IP_\alpha}$ and $\braket{ZP_\alpha}$ at $\mathrm{Im}(\alpha) = 0$ show the increasing separation of the coherent state superpositions, whereas $\braket{XP_\alpha}$ and $\braket{YP_\alpha}$ at $\mathrm{Re}(\alpha) = 0$ reveal the interference fringe oscillations dependence on cat state size. (b) Single cuts at $\beta = \sqrt{3}$ show single-shot correlations (crosses) with their ideal trends (solid line). Using just individual measurement settings (circled), joint observables such as $\{II_c,XX_c,YY_c,ZZ_c\} $ of the qubit-cavity state can be determined.
}
   \label{fig:3}
\end{SCfigure*}
The number of measurement settings required to perform cavity state tomography can be resource intensive. Restricting to an encoded qubit subspace, only four values of the cavity Wigner function $W(\alpha)$ are required to reconstruct the state, known as a direct fidelity estimation (DFE)~\cite{Flammia:2011dfe,daSilva:2011ej}. For large cat states $|\braket{\beta | {-\beta}}|^2 \ll 1$, the encoded state observables map to cavity observables as:
\begin{align}
X_c &   =   {P_{0}} &   {I_c} &   =     {P_{\beta}} + {P_{{-\beta}}} \\ \nonumber
{Y_c} &   =   {P_{\frac{j\pi}{8\beta}}}  &   {Z_c} &   =     {P_{\beta}} - {P_{{-\beta}}}
\end{align}
where $\{I_c,X_c,Y_c,Z_c\}$ form the Pauli set for the encoded qubit state in the cavity (see methods). Cuts in the joint Wigner function~(Fig.~\ref{fig:3}) show these observables and their correlations to the qubit as a function of cat state size. As the superposition state is made larger, interference fringe oscillations increase while fringe amplitude decreases due to photon loss. For a state $\ket{\psi_\B}$ with $|\beta| = \sqrt{3}$, we estimate a direct fidelity $\mathcal{F}_\mathrm{DFE} = \frac{1}{4}(\braket{II_c} + \braket{XX_c} - \braket{YY_c} + \braket{ZZ_c}) = (72 \pm 2)\%$ putting a fidelity bound on the target state with no corrections for visibility. This estimate is related to the benchmarks reported above $\mathcal{F}_\mathrm{DFE} \approx \mathcal{V} \times \mathcal{F}$ and far surpasses the 50\% threshold for a classically correlated state. This indicates both high fidelity state-preparation and measurement, and demonstrates that strong correlations are directly detectable using joint Wigner tomography.

\begin{figure*}[t]
\centering
\includegraphics[scale = 0.8]{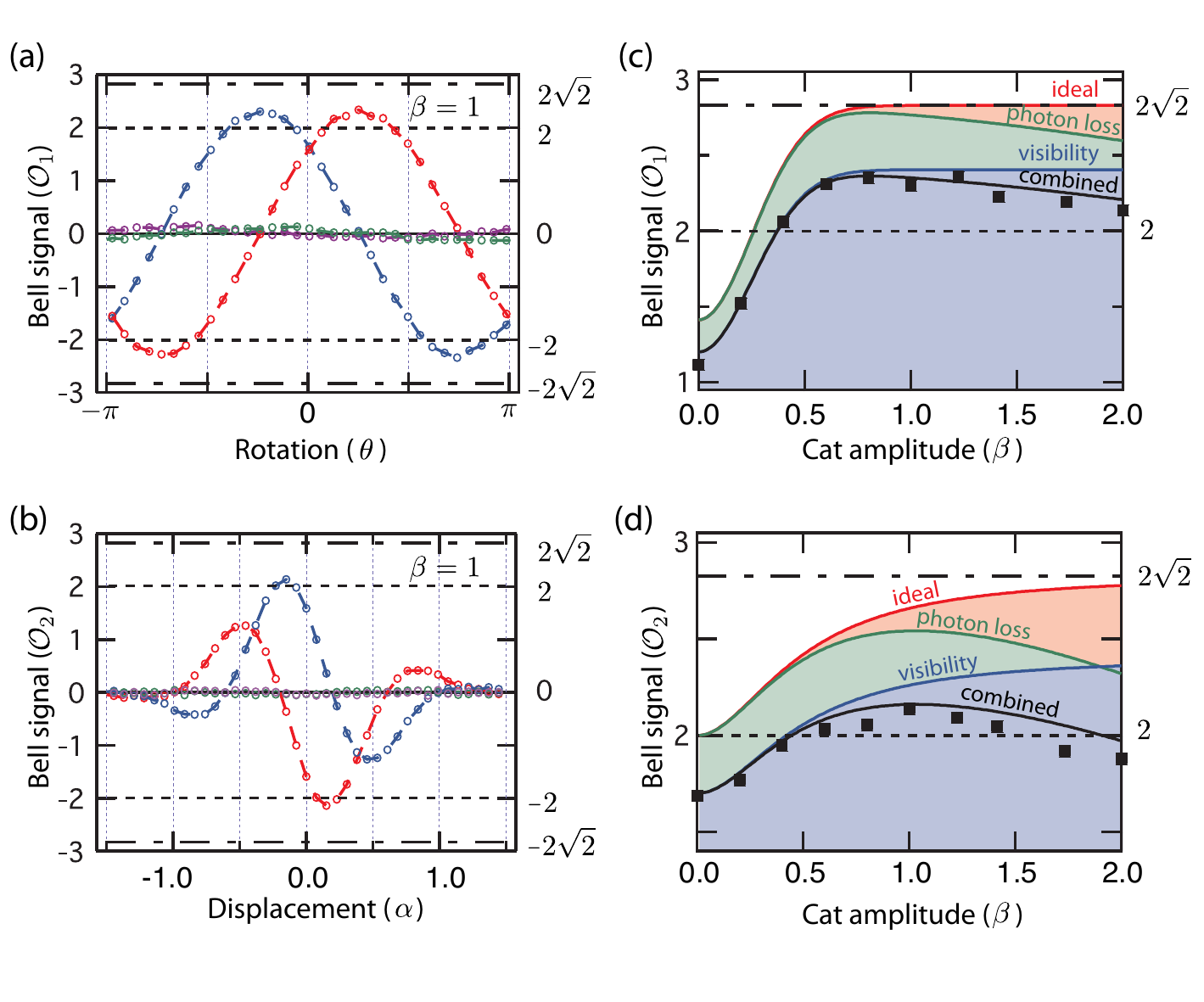}
\caption{\textbf{Bell tests with a cat state.}
 (a) Using correlations between qubit state observables $X(\theta)$ and $Z(\theta)$ and the encoded state observables $Z_c$, $X_c$; a CHSH Bell test $\mathcal{O} = \braket{AA_c} + \braket{AB_c} - \braket{BA_c} + \braket{BB_c}$  is performed as a function of qubit detector angle $\theta$. (b) Furthermore, we observe a violation with an additional Bell test using qubit observables $X$, $Y$ and cavity state observables $X_c(\alpha)$, $Y_c(\alpha)$ where $\alpha$ corresponds to a tomography displacement amplitude serving as a rotation of the effective cavity detector angle. (c-d)~Both tests are performed for different cat state amplitudes $\beta$ and show the dependence of the entangled state with photon loss and detector visibility. Squares represent measured values with height denoting their statistical uncertainty. Solid lines describe the predicted trends given the measured cavity decay rate and detection visibility. While the ideal behaviour (red) for an entangled state approaches $\mathcal{O} = 2\sqrt{2}$, photon loss (green), detector visibility (blue), and their combined effects (black) will ultimately limit the maximum Bell signal achieved.
}
   \label{fig:4}
\end{figure*}
To place a stricter bound on observed entanglement, we perform a CHSH Bell test on the measured state. Although proposed to investigate local hidden variable theory, the Bell test also serves to benchmark the performance of a quantum system that creates and measures entangled states~\cite{Ansmann:2009eq,Pfaff:2012gy,Chow:2010dd}. Classical theory dictates that the sum of four correlations will be bounded such that:
\begin{equation}
-2 \leq \mathcal{O} = \braket{AA_c} + \braket{AB_c} - \braket{BA_c} + \braket{BB_c} \leq 2
\end{equation}
where in this experiment $A,B$ are two qubit observables and $A_c, B_c$ are two cavity observables. We perform two Bell tests (Fig.~\ref{fig:4}) with correlations taken shot-by-shot with no post-selection or compensation for detector inefficiencies. In the first, we take observables $X(\theta),Z(\theta),X_c, Z_c$ and sweep both qubit detector angle $\theta$ (see methods) and cat state amplitude $\beta$. We observe a Bell signal with a maximal violation $\mathcal{O}_1 = 2.30\pm 0.04$ at $\theta = -\frac{\pi}{4}$ for $\beta = 1$. In the second Bell test, we follow a scheme similar to Ref.~\cite{Park:2012iw} and choose observables $X,Y,X_c(\alpha),Y_c(\alpha)$ where $\alpha$ is a displacement amplitude corresponding to a rotation of the encoded cavity state detector (see methods) and observe a maximal violation $\mathcal{O}_2 = 2.14\pm0.03$ for $\beta = 1$. As predicted, a lower Bell signal is observed in the second test due to its greater sensitivity to photon loss, yet in both tests two regimes are evident. For small cat state amplitudes, the initial Bell signal is limited by the non-orthogonality of the coherent state superpositions (see methods), while for large displacements the system's sensitivity to photon loss results in a reduction of the Bell signal. Larger, more distinguishable states quickly devolve into a classical mixture due to the onset of decoherence, corresponding to the resolution of Schr\"{o}dinger's thought experiment. However, for intermediate cat state sizes, we violate Bell's inequality beyond the statistical uncertainties in both tests.

In this letter, we have demonstrated and quantified the entanglement between an artificial atom and a cat state in a cavity mode. We determine the entangled state using sequential detection with high-fidelity state measurement and real-time feedback on the quantum state. We benchmark the capabilities of this detection scheme with direct fidelity estimation and Bell test witnesses, which both reveal non-classical correlations of our system. This work demonstrates the viability of using and measuring redundantly encoded states in multi-level systems~\cite{Gottesman:2001jb}. This implementation provides a vital resource for quantum state tomography and quantum process tomography of continuous-variable systems and creates a platform for measurement based quantum computation and quantum error correction using superconducting cavity resonators~\cite{Mirrahimi:2014js}. Finally, these features are directly extendible to multi-cavity systems which will require entanglement detection between continuous variable degrees of freedom and entanglement distribution of complex oscillator states.

\subsection*{Acknowledgments}
We thank W. Pfaff, A. Narla, and R. Heeres for discussions. This research was supported by the National Science Foundation (NSF) (PHY-1309996), the Multidisciplinary University Research Initiatives program (MURI) through the Air Force Office of Scientific Research (FA9550-14-1-0052), and the U.S. Army Research Office (W911NF-14-1-0011). Facilities use was supported by the Yale Institute for Nanoscience and Quantum Engineering (YINQE) and the NSF (MRSECDMR 1119826).

\bibliographystyle{naturemag}
\bibliography{ref}

\twocolumn[
  \begin{@twocolumnfalse}
\begin{center}
\textbf{\Large Supplemental Materials:\\ \large Violating Bell's inequality with an artificial atom and a cat state in a cavity}
\end{center}
  \end{@twocolumnfalse}
]
\setcounter{equation}{0}
\setcounter{figure}{0}
\setcounter{table}{0}
\setcounter{page}{1}
\makeatletter
\renewcommand{\theequation}{S\arabic{equation}}
\renewcommand{\thefigure}{S\arabic{figure}}
\renewcommand{\thetable}{S\@arabic\c@table}

\section{Materials and methods}
        \label{sec:methods}
\bss{Measurement setup: }Experiments are performed in a cryogen-free dilution refrigerator at a base temperature of $\sim10$~mK.  Our output signal amplification chain consists of two stages. A Josephson bifurcation amplifier (JBA) \cite{Vijay:2009ko} operating in a double-pumping configuration \cite{Kamal:2009vx,Murch:2013bd} serves as the first stage, which is followed by a high electron mobility transistor (HEMT) amplifier.

Fabrication techniques of the transmon qubit and the design of storage and readout resonators follow the methods described in \cite{Vlastakis:2013ju}. The refrigerator wiring (see Fig.~\ref{fig:wiring}), including the filters and attenuators used, are similar to that of \cite{Sun:2014ha}, but with the addition of a feedback system, the details of which are discussed in a following section.
\begin{figure*}[h]
    \includegraphics[scale=0.7]{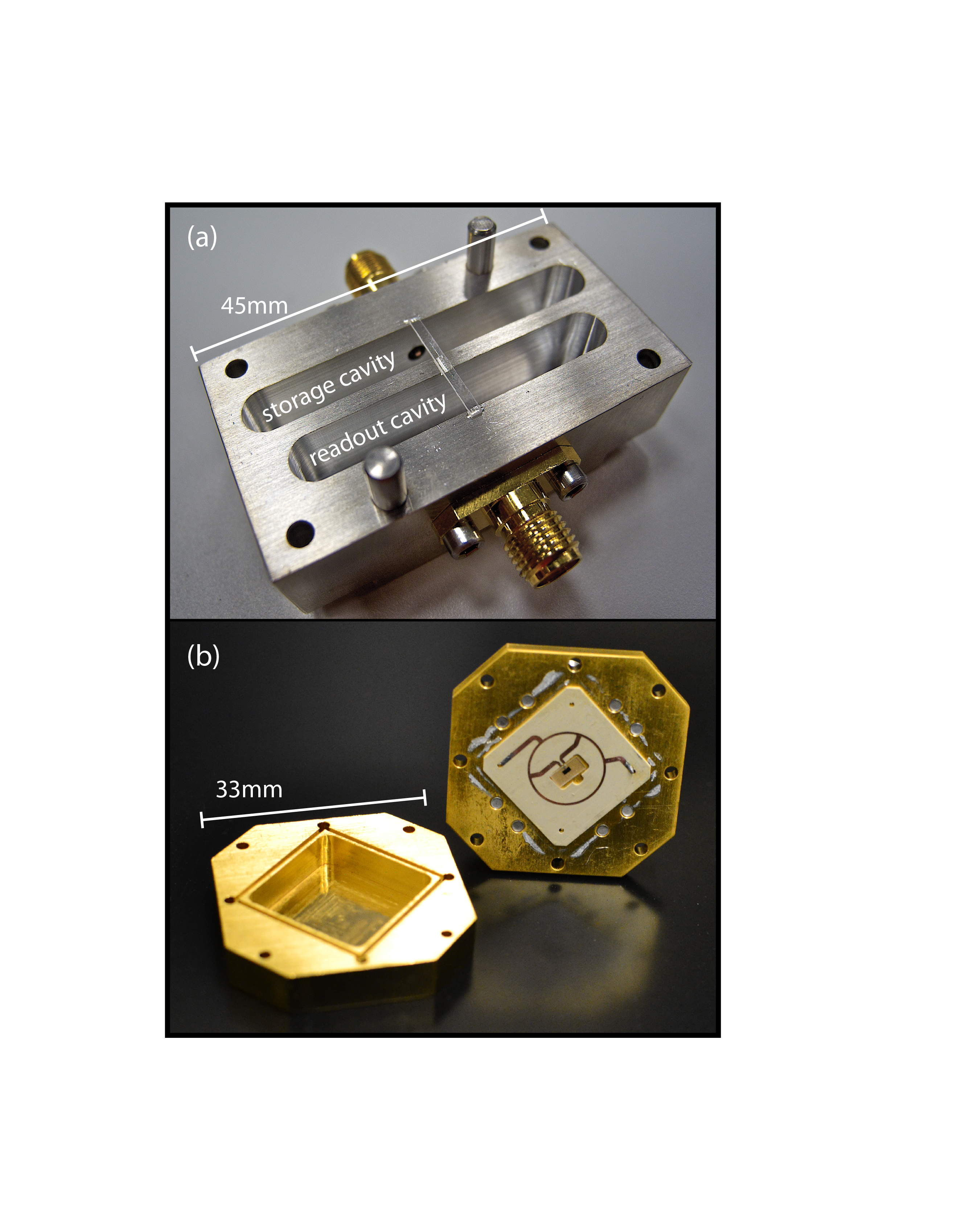}
        \centering
    \caption{\textbf{Photograph of device and amplifier: } (a) One half of the 3D circuit QED device shows both the readout and storage cavities. Strongly coupled to each cavity is a single vertical transmon. (b) High-fidelity measurements are achieved with near quantum-limited amplification provided by a Josephson bifurcation amplifier. Shown is the chip and sample holder for this device.}
        \label{fig:device}
\end{figure*}
\begin{figure*}[h]
    \includegraphics{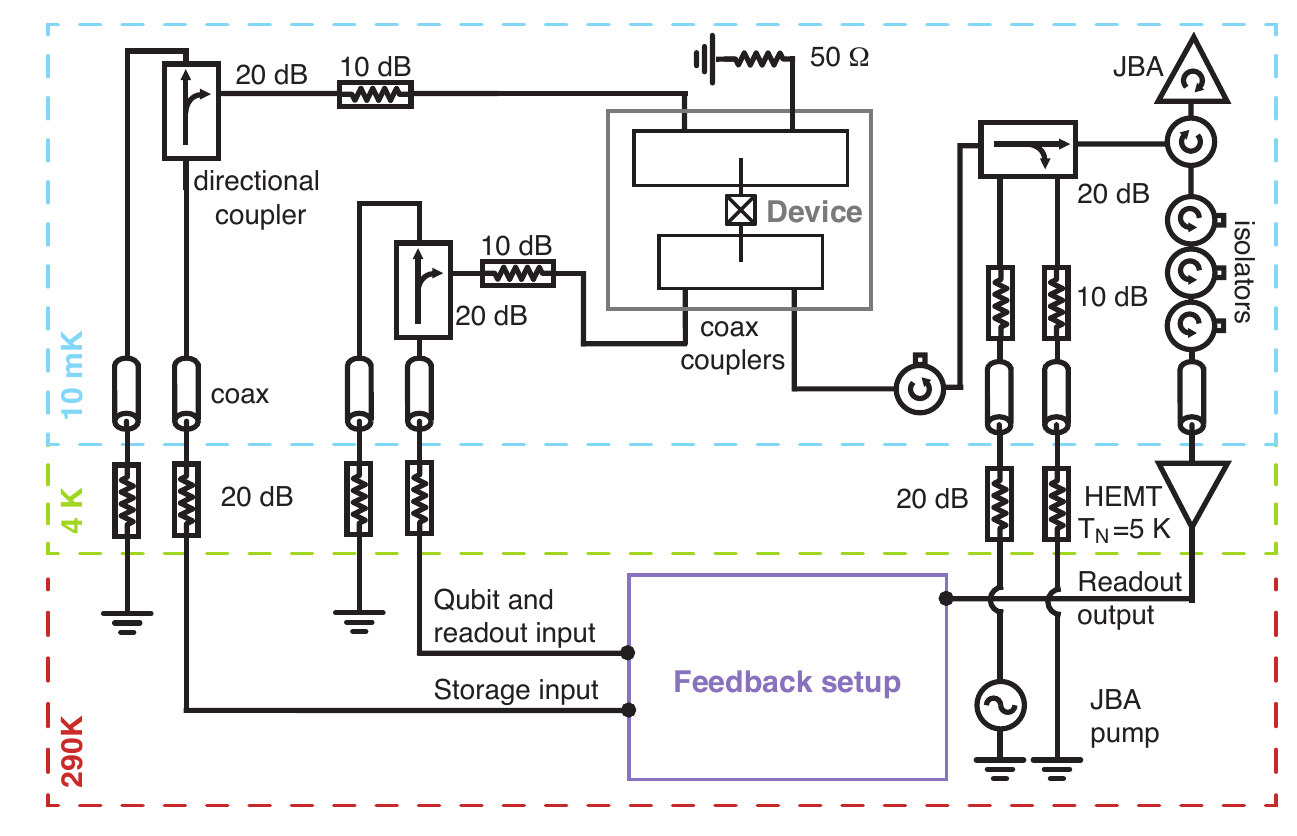}
        \centering
    \caption{\textbf{Experiment Schematic}}
        \label{fig:wiring}
\end{figure*}
\begin{figure*}[h]
\caption{\textbf{Feedback inset: }The feedback setup uses two input-output (I/O) boards for qubit and storage resonator control and one arbitrary waveform generator (AWG) for readout resonator control.  All have a dedicated microwave generator and mixer for amplitude and phase modulation.  Each I/O board has five main components: 1) a digital-to-analog converter (DAC) for pulse generation; 2) digital outputs serving as marker channels; 3) an analog-to-digital converter (ADC) that samples input signals; 4) an FPGA that demodulates the signals from the ADC and based on predefined thresholds determines the measured qubit state, $\ket{g}$ or $\ket{e}$ to generate pulses; and 5) a PCIe connection that transfers FPGA data to a computer (PC) for analysis.  In this setup the top I/O board serves as the master, which accepts the readout signal, returns qubit state information, and using digital output signals, triggers the AWG and the second I/O card given a particular qubit measurement result.}
\centering
\includegraphics{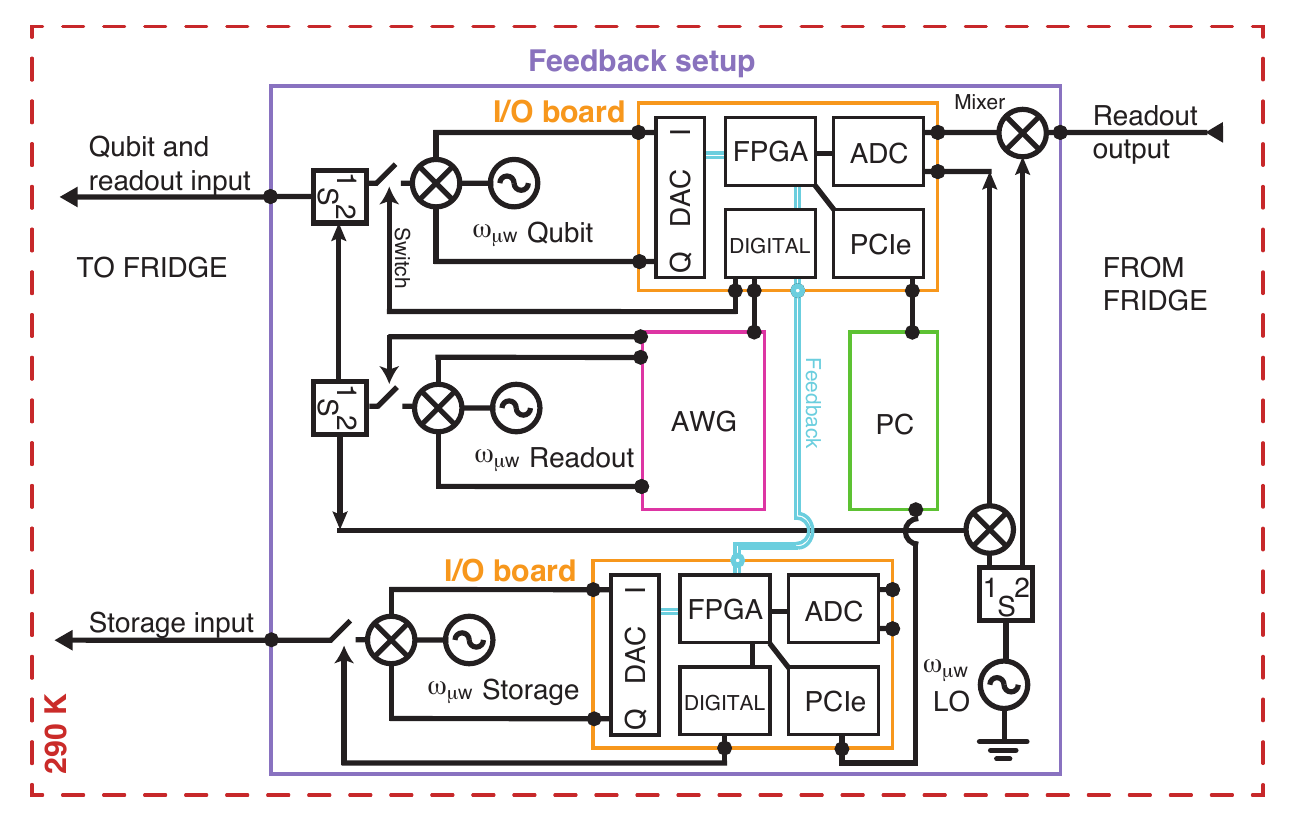}
        \label{fig:inset}
\end{figure*}

\bs{Qubit-cavity parameters: } The two-cavity, single-qubit system is well described by the approximate dispersive Hamiltonian:
\begin{align}
H/\hbar &= \omega_s a_s^\dagger a_s + \omega_r a_r^\dagger a_r + \omega_q b^\dagger b \\ \nonumber
& - \tfrac{K_s}{2}{a_s^\dagger}^2 {a_s}^2 - \tfrac{K_r}{2}{a_r^\dagger}^2 {a_r}^2 - \tfrac{K_q}{2}{b^\dagger}^2 {b}^2 \\ \nonumber
& - \chi_{qs}  a_s^\dagger a_s b^\dagger b - \chi_{qr} a_r^\dagger a_r b^\dagger b - \chi_{rs} a_s^\dagger a_s a_r^\dagger a_r
\end{align}
Where $\omega_{s,r,q}$ are the storage, readout, and qubit transition frequencies, $a_s, a_r, b$ are the associated ladder operators, and $K,~\chi$ are the modal anharmonicities and dispersive shifts respectively. Table~\ref{tab:params} details the Hamiltonian parameters of our system.  The resonant frequency of the readout resonator $\omega_r/2\pi$ is determined by transmission spectroscopy.  The qubit frequency $\omega_s/2\pi$ and storage cavity frequencies $\omega_q/2\pi$ are found using two-tone spectroscopy.

Qubit anharmonicity $K_q$ is measured using two-tone spectroscopy to observe the $0-2$ two-photon transition \cite{Paik:2011hd}.  Storage cavity anharmonicity $K_s$ is determined by displacing the cavity with a coherent state and observing its time evolution with Wigner tomography.  The resulting dynamics are characterized by state reconstruction and $K_s$ is observed by the state's quadratic dependence of phase on photon number. Finally, we predict the readout cavity anharmonicity $K_r$ using its approximate dependence on the measured values of $K_q$ and the qubit-readout dispersive shift $\chi_{qr}$ \cite{Nigg:2012jja}.

The dispersive shift between the qubit and the readout resonator $\chi_{qr}$ is found by taking the difference in frequency between the readout resonance when the qubit is in the ground and excited state.  The dispersive shift between the qubit and the storage resonator $\chi_{qs}$ is found using two methods: photon number dependent qubit spectroscopy \cite{Schuster:2007ki}, and observing qubit state revival using Ramsey interferometry \cite{Vlastakis:2013ju}.  Finally, $\chi_{rs}$ is predicted using its approximate relationship between $K_s$ and $K_r$ \cite{Nigg:2012jja}.

\bs{Lifetimes and thermal populations: }The lifetime of the storage cavity is determined by displacing to a coherent state, waiting a variable length of time, and then applying a qubit rotation conditioned on zero photons in the storage cavity. This allows a measurement of the time-dependent overlap of the cavity state with its ground state $\ket{0}$ dependent on time.  The lifetime of the readout cavity is found from its line-width.  The thermal population of the qubit is determined from a histogram of one million single-shot measurements of the qubit thermal state, where the signal-to-noise ratio provided by the JBA allows discrimination between $\ket{g}$ and all states not $\ket{g}$.  The thermal population of the storage cavity is found by taking the difference between parity measurements of the thermal and vacuum states of the cavity.  A vacuum state is prepared by first performing two parity measurements on the thermal state and then post-selecting such that all results give even parity, projecting the thermal state onto $\ket{0}$.  Finally, the known thermal population of the readout cavity is bounded by the dephasing rate $\Gamma_{\phi}$ of the qubit: $\Gamma_{\phi}$=$\bar{n}_{th}\kappa$, where $\bar{n}_{th}$ is the readout cavity's thermal occupation and $\kappa$ is the readout single-photon decay rate~\cite{Sears:2012cm}.
\begin{figure*}
\begin{floatrow}
\capbtabbox[3in]{%
\begin{tabular}{cc}

\hline\hline

 Term & Measured  \\
      & (Prediction) \\

 \hline

 $\omega_q/2\pi$ & 5.7651 GHz \\
 $\omega_s/2\pi$ & 7.2164 GHz \\
 $\omega_r/2\pi$ & 8.1740 GHz \\

 \hline

 $K_q/2\pi$ & 240 MHz \\
 $K_s/2\pi$ & 1.5 kHz \\
 $K_r/2\pi$ &   (2 kHz)\\ \hline

 $\chi_{qs}/2\pi$ & 1.43 MHz \\
 $\chi_{qr}/2\pi$ & 1 MHz \\
 $\chi_{rs}/2\pi$ & (1.7 kHz)\\

 \hline

\end{tabular}
\label{tab:params}
}
{%
    \caption{\textbf{Hamiltonian parameters} }%
}
\capbtabbox[3in]{%
\begin{tabular}{cccc}
\hline\hline
 & Qubit & Storage & Readout \\ \hline
$T_1$ & $10\mu$s & - & -\\
$T_2$& $10\mu$s & - & -\\
$\tau_\mathrm{cav}$ & - & $55\mu$s & $30$ns\\
\hline
ground state (\%)& $90\%$ & $>98\%$ & $>99.8\%$ \\
\hline

\end{tabular}
\label{tab:coherence}
}
{%
  \caption{\textbf{Coherence and thermal properties} }%
}
\end{floatrow}
\end{figure*}
\begin{figure*}[h]
\centering
\includegraphics[]{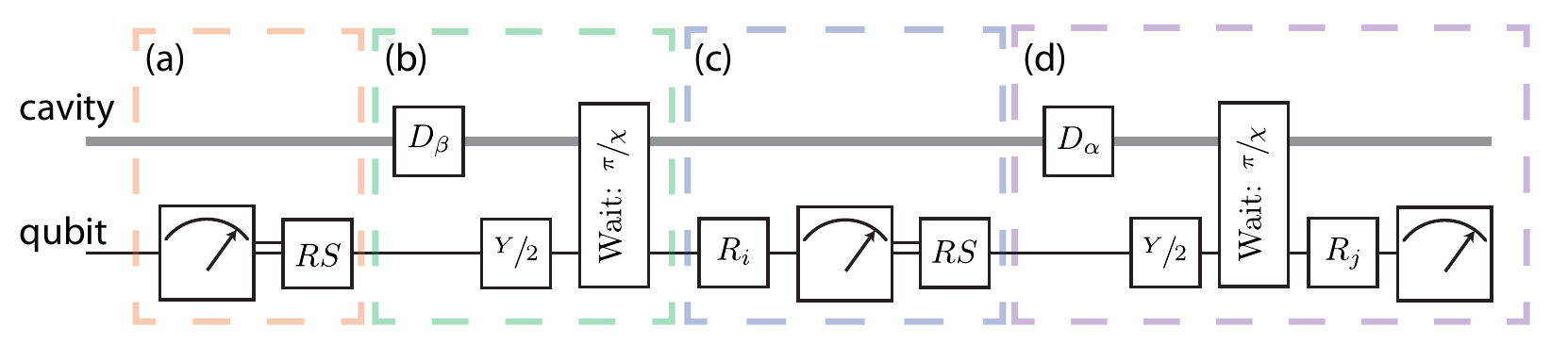}
\caption{\textbf{Full quantum circuit diagram}
Each experiment is split into four components. (a) First, the system is initialized. The qubit state is measured and a qubit pulse $RS = \{ R^\pi_{\hat y} \textrm{ or } \mathbbm{1}\}$ is applied to reset the qubit to $\ket{g}$. (b) Second, the entangled state is created with a cavity displacement $D_\beta$ and a qubit rotation $Y/2 = R^{\pi/2}_{\hat y}$ followed by a $\frac{\pi}{\chi}$ waiting time to produce the entangled state $\ket{\psi_\mathrm{B}} = \frac{1}{\sqrt{2}}(\ket{g,\beta} + \ket{e, -\beta})$. (c) Following preparation, a qubit state detection is performed with a pre-rotation $R_i$ (see Table~S3), a measurement, and a qubit reset $RS$. Finally, we perform a cavity state measurement using Ramsey interferometry where $R_j = \pm Y/2$ combined with an initial pre-displacement $D_\alpha$. This maps $P_\alpha$ to the qubit state which is read out with a subsequent qubit measurement. Correlations are reported as the product of detection events between measurements in (c) and (d).
}
   \label{fig:circui_full}
\end{figure*}

\bs{Measurement fidelities: }We define singleshot measurement fidelity as $F_q= \frac{P(g|g)+P(e|e)}{2}$, where $P(g|g)$ and $P(e|e)$ are the probabilities to get $\ket{g}$ ($\ket{e}$) knowing that we start with $\ket{g}$ ($\ket{e}$).  The state $\ket{g}$ is prepared through purification of the qubit thermal state with realtime feedback (see the following section).  Given a preparation of $\ket{g}$, we have a $98.5\%$ chance of measuring $\ket{g}$ again ($P(g|g)=0.985$).  Likewise, we find $P(e|e)=0.975$ by preparing $\ket{g}$ and rotating the state to $\ket{e}$.  This gives a single-shot measurement fidelity of $F_q = 98\%$.  We find our cavity parity measurement fidelity by purifying the storage cavity thermal state into $\ket{0}$ then performing one of two kinds of parity measurement (see Fig.~\ref{tab:settings}). We report a parity measurement fidelity for $n=0$ photons as $F_c = \frac{P(g|E_1)+P(e|E_2)}{2}= 95.5\%$, where $P(g|E_1)$ ($P(e|E_2)$) is the probability to measure $\ket{g}$ ($\ket{e}$) given that the parity is even for each of the two measurement settings. We expect $F_c$ to decrease with increasing numbers of photons in the cavity due to single photon loss during the measurement sequence.

Directly from these readout fidelities, the estimated visibility \cite{Kofman:2008gq} for correlated observables $\mathcal{V}_\mathrm{est} = (2F_q - 1)(2F_c - 1) = 87\%$. This allows us to predict the maximum Bell violation possible given only measurement inefficiencies $\mathcal{O}_\mathrm{max} = 2\sqrt{2}\mathcal{V}_\mathrm{est} = 2.47$. In practice, $\mathcal{V}$ is directly related to the contrast of the joint Wigner function (see Sec.~\ref{sec:tomo}) which we measure to be $85\%$. This discrepancy is due to qubit decoherence, which is studied further in Sec.~\ref{sec:errors} and puts a more conservative estimate for the maximum Bell violation achievable: $\mathcal{O}_\mathrm{max} = 2\sqrt{2}\mathcal{V} = 2.40$.

\bs{I/O control parameters: }As shown in Fig.~\ref{fig:inset}, we employ a field-programmable gate array (FPGA) in order to implement an active feedback scheme. We use an X6-1000M board from Innovative Integration which contains two 1 GS/s ADCs, two 1 GS/s DAC channels, and digital inputs/outputs all controlled by a Xilinx VIRTEX-6 FPGA loaded with custom logic.  We synchronize two such boards in a master/slave configuration to have IQ control of both the qubit/storage cavity. IQ control over the readout cavity is performed with a Tektronix AWG, which is triggered by the master board.  The readout and reference signals are routed to the ADCs on the master board, where after the FPGA demodulates the signal and decides whether the qubit is in $\ket{g}$ or $\ket{e}$. The feedback latency of the FPGA logic (last in, first out LIFO) is 320 ns. Additional delay for active feedback include cable delay ($\sim 100~\textrm{ns}$) and readout pulse length with resonator decay time ($320~\mathrm{ns}$).  Thus, in total the qubit waits $\tau_{\mathrm{wait}} \sim 740~\mathrm{ns}$ between the time photons first enter the readout resonator and the time at which the feedback pulse resets the qubit.

\bs{Implementations of feedback: }Feedback is used three times during a single iteration of the experiment.  Prior to the state preparation (Fig.~\ref{fig:circui_full}), we purify the qubit state to $\ket{g}$ by measuring the qubit and applying a rotation $R^\pi_{\hat y}$ if measured in $\ket{e}$.  We succeed in preparing $\ket{g}$ with a probability of $99\%$.  Secondly, when performing qubit tomography we reset the qubit to $\ket{g}$ if it is measured to be in $\ket{e}$.  Since we must wait $\tau_{\mathrm{wait}}$ before feedback can be applied, the cavity state will acquire an additional phase $\chi_{qs}\tau_{\mathrm{wait}}$ if the qubit is in $\ket{e}$.  In this case, in addition to reseting the qubit, the FPGA applies an equivalent phase shift on the subsequent Wigner tomography pulse. This feedback implementation does not close the `locality' loophole for a CHSH Bell test and therefore cannot be used to test local realism.


\section{Random and systematic errors}

\label{sec:errors}

\bss{Gaussian error statistics: }We perform single-shot measurements that are discriminated into binary results and report measured observables taking the mean of $N$ experimental outcomes. All mean values are reported with $N>4000$ measurements and the highest measurement fidelity for any joint observable is $\mathcal{F} = (\mathcal{V}+1)/2 = 0.93$ such that $ \textrm{min}[N\mathcal{F}, N(1-\mathcal{F})]> 300 \gg 1$. From this, we can approximate the mean value of all measured observables to follow a Gaussian distribution with standard deviation $\frac{1}{\sqrt{N}}$ (Fig.~\ref{fig:residuals}).
%
%

\bs{Detector cross-talk:} The sequential detection protocol in this experiment uses the same detector to perform first a qubit  measurement followed by a cavity measurement. To minimize unwanted systematic errors due to detector cross-talk  between measurements, we perform each experiment under four detector setting permutations. Two settings are used for the qubit measurement: a pre-rotation which maps a qubit eigenstate $\ket{\pm}$ to detector values $\pm M^q_1$ and another which maps $\ket{\pm}$ to $\mp M^q_1$. Two settings are used for the cavity measurement: a Ramsey experiment which maps a cavity eigenstate $\ket{\pm}$ to detector values $\pm M^c_2$ and another which maps $\ket{\pm}$ to $\mp M^c_2$ (Tab.~S3). Each detector setting is performed an equal number of times and results are combined to remove unwanted correlations between detector readings and measured quantum observables. See Sec.~\ref{sec:bell} for an analysis on the effects of these detector settings on a Bell test.

\begin{figure*}
  \includegraphics[scale=0.6]{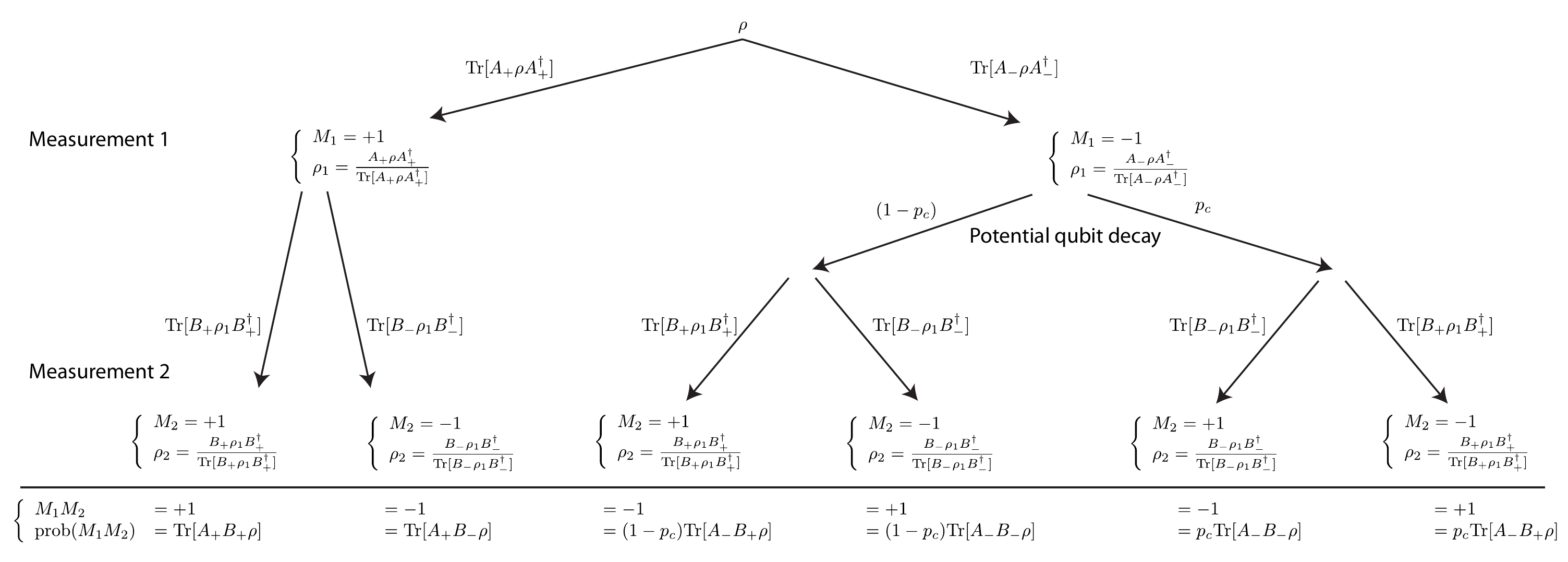}
  \caption{\textbf{Measurement trajectories given qubit decoherence.}
We can model the behaviour of qubit decoherence in a single measurement trajectory. Qubit decay (which occurs with a probability $p$) can lead to an improper initialisation of the second detection and in turn produces an incorrect measurement result. This form of detector cross-talk can lead to a reduction in visibility and potential systematic offset of the measured qubit-cavity observable: $\braket{AB} \to (1 - p_c) \braket{AB} - p_c\braket{B}$ (see section 2).
}%
    \label{fig:tree}
\end{figure*}

\begin{figure*}[h]
\begin{floatrow}
\ffigbox{%
  \includegraphics[]{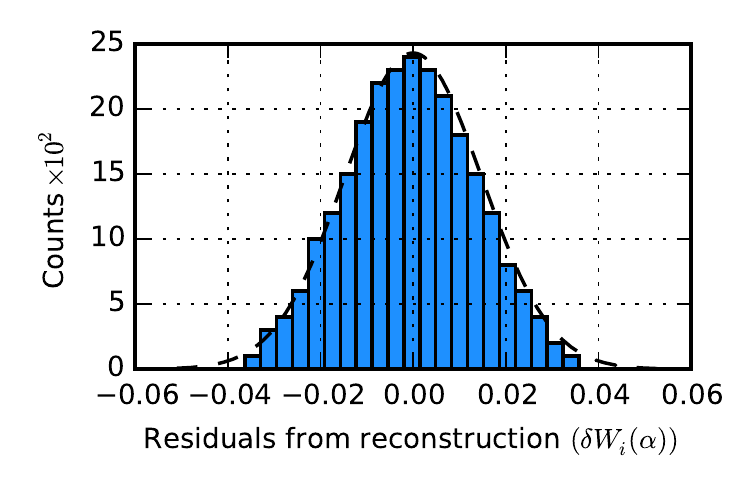}
}{%
  \caption{\textbf{Histogram of reconstruction residuals.}
Plotted are the residuals corresponding to the density matrix reconstruction of the Bell cat state shown in Fig.~2 of the main text. This Histogram shows the distribution of the $~25,000$ residuals from the joint Wigner function which gives a Gaussian distribution (mean value $\mu = 7.0\times10^{-4}$, standard deviation $\sigma = 0.015$), which agree with our expectation for statistical error $\sigma_\mathrm{est} = \frac{1}{\sqrt{N}} \approx 0.015$.
}%
\label{fig:residuals}
}
\capbtabbox[3in]{%
$\begin{array}{ll|ll}
\hline \hline
\multicolumn{2}{c}{\textrm{Qubit}} & \multicolumn{2}{c}{\textrm{Cavity}} \\
\cline{1-2} \cline{3-4}
R_i                     & M_1   & R_j      & M_2  \\
\hline
\mathbbm{1}             & +Z    & R^{\pi/2}_{\hat y}      & +P_\alpha \\
R^\pi_{\hat y}          & -Z    & R^{-\pi/2}_{\hat y}     & -P_\alpha  \\
R^{\pi/2}_{\hat y}      & +X    &                &    \\
R^{-\pi/2}_{\hat y}     & -X    &                &    \\
R^{-\pi/2}_{\hat x}     & +Y    &                &    \\
R^{\pi/2}_{\hat x}      & -Y    &                &    \\
\hline
\end{array}$
}
{%
  \caption{\textbf{Table of measurement operators.} As shown in Fig.~\ref{fig:circui_full}, pre-rotations before qubit and cavity state measurements determine the measured observable. Shown are the different pre-rotations used and the corresponding measurement operator.}%
}
\label{tab:settings}
\end{floatrow}
\end{figure*}

The dominant form of cross-talk for this experiment is due to qubit state decoherence between measurements. To realize the cavity state measurement, the qubit must be initialized in $\ket{g}$, which we perform using active feedback. Qubit decay can occur during this reset process causing an incorrect initialisation for cavity state detection. We can model this error by observing the possible trajectories of each measurement outcome (Fig.~\ref{fig:tree}). This modifies the average measurement of the observable $AB$ where $A,B$ are qubit and cavity operators that can be decomposed into qubit projectors $AB = (A_+ - A_-)B$, where $A_+ + A_- = I$. Due to qubit decay, the measured value $\mean{A_+B}$ will be modified to $(1-2p_c)\mean{A_+B}$ where $p_c$ is the probability of qubit decay in the time between the first measurement and the feedback rotation. This relation changes the measurement into:
\begin{align}
\braket{AB}  \to &(1 - 2p_c)\braket{A_+B} - \braket{A_-B} \\ \nonumber
     =  &(1 - p_c)\braket{A_+B - A_-B} - p_c\braket{A_+B + A_-B} \\ \nonumber
     =  &(1 - p_c)\braket{AB} - p_c\braket{B}
\end{align}
For measuring $B = X_c, Y_c, Z_c$ of the Bell-cat state $\ket{\psi_c}$, we expect $\mean{B} = 0$, which gives merely a reduction in the visibility of the observable $\braket{AB}$ by a factor$(1 - p_c)$ without systematic offsets. We estimate in this experiment that $p_c = 1 - e^{-\frac{\tau_\mathrm{wait}}{T_1} } \approx 0.06$. With this justification we can predict the additional loss in visibility $\mathcal{V}$ mentioned in the previous section which gives a visibility $\mathcal{V}_\mathrm{pred} = (1 - p_c)\mathcal{V} = 82\%$. The experimentally obtained visibility $\mathcal{V}$ is $85\%$; we believe the discrepancy between predicted and measured values is due to an overestimate in the time the qubit is susceptible to energy decay during measurement.
%

%
%
\bs{Tomography rotation errors:} We observe systematic effects attributed to an amplitude error using the $R^\pi_{\hat y}$ operation for pre-rotations used in qubit state tomography. A arbitrary pre-rotation $R^{\theta,\phi} = e^{\frac{i\pi\theta}{2}(\sigma_y\cos\phi + \sigma_x\sin\phi)}$ transforms a qubit measurement along the $\hat Z$ axis:
\begin{equation}
\label{eq:prerot}
\hat Z \to \cos{\theta}\hat Z + \sin{\theta}\cos{\phi} \hat X + \sin{\theta}\sin{\phi} \hat Y
\end{equation}
If $\theta \neq 0 \textrm{ or } \pi$, a systematic offset can occur. This is observed in Fig.~\ref{fig:pauli_set} where the contrast of $\langle ZZ_c \rangle$ is reduced and a residual offset in $\langle ZX_c\rangle$ and $\langle ZY_c \rangle$ respectively are produced. From Eq.~\ref{eq:prerot}, we predict the following relationship:
\begin{align}
\braket{ZZ_c}\tan{\theta} = \sqrt{ \braket{ZX_c}^2 + \braket{ZY_c}^2 }
\end{align}
From measurements, we can approximate the fractional amplitude rotation error $\delta \theta = 4.2 \%$. This under-rotation is due to a photon-dependence of the calibrated pulse amplitude. Mitigating this error susceptibility is being explored with composite pulses in future experiments.


\section{State tomography}
        \label{sec:tomo}
\bss{Calculating observables:} We can represent the density matrix in the excitation number basis:
\begin{equation}
\rho = \sum_{i,j = 0}^1 \sum_{n,m = 0}^N \rho^{nm}_{ij} \ket{i}\bra{j} \otimes \ket{n}\bra{m}
\end{equation}
where $\rho^{nm}_{ij}$ are elements of qubit/cavity density matrix and $\ket{i,j}$ is the qubit state in the excitation basis and $\ket{n,m}$ is the cavity state in the excitation (photon number) basis. From a density operator, one can calculate an observable of the combined system by determining the product of observables from each individual system:
\begin{equation}
\langle A  B \rangle = \mathrm{Tr} \left[  A  B  \rho \right]
\end{equation}
where $A$, $B$ are operators for the qubit and cavity respectively.

In the joint Wigner function, the qubit basis is the Pauli set $ \sigma_i =  \left \{  I,  \sigma_x,  \sigma_y,  \sigma_z \right \}$. For the cavity mode, we choose the displaced photon parity operator $P_\alpha = D_\alpha P D_\alpha^\dagger$ that corresponds to a single point in the cavity state Wigner function. For a truncated Hilbert space (in this experiment $N_\mathrm{max} = 12$ ) and a displacement grid of $\alpha_\mathrm{max,min} = \pm 3.4$ with step size $\Delta \alpha = 0.085$, this measured Wigner function represents an over-complete set of measurements for the cavity mode.
The joint Wigner function $W_i(\alpha) = \frac{2}{\pi}\langle \sigma_i  P_\alpha \rangle$ is constructed directly from experimental measurements.

A qubit operator $A$ can be written in the Pauli basis $ A = \sum_i A_i  \sigma_i$ where $A_i = \mathrm{Tr}[ A  \sigma_i]$ and a bounded cavity observable (see \cite{Cahill:1969jg} for details) can be represented in continuous-variable basis $ B = \frac{1}{\pi} \int B(\alpha)  P_\alpha \mathrm{d}^2 \alpha$ where $B(\alpha) = \mathrm{Tr}[ B P_\alpha ]$. Finally, the composite qubit-cavity density matrix can be written as:
\begin{equation}
\rho = \pi \sum_i \int W_i(\alpha) \sigma_i P_\alpha \mathrm{d}^2 \alpha
\end{equation}
Note that for separable states $\rho = \rho_q \otimes \rho_c$, this relation can be split up into their respective discrete and continuous components:
\begin{equation}
 \rho = \frac{1}{2} \sum_i \mathrm{Tr}[ \rho_q \sigma_i] \sigma_i \otimes 2 \pi \int \frac{2}{\pi}\mathrm{Tr[ \rho_c P_\alpha]} P_\alpha \mathrm{d}^2 \alpha
\end{equation}
For any state $\rho$, we can write the mean value of an observable for the combined system with the following relation:
\begin{align}
    \label{eq:overlap}
\langle  A  B \rangle &= \mathrm{Tr}\left[ A  B  \rho \right] \nonumber \\
&= \mathrm{Tr} \left[ \sum_{i,j}\int A_i B(\alpha) W_j(\alpha')  \sigma_i  \sigma_j  P_\alpha  P_{\alpha'} \mathrm{d}^2 \alpha \mathrm{d}^2 \alpha' \right]
\end{align}
Using the following operator rules $\mathrm{Tr}[ \sigma_i  \sigma_j] = \delta_{ij}$ and $\mathrm{Tr}[ P_\alpha  P_{\alpha'}] = \delta^2(\alpha -\alpha')$ we can simplify Eq.~\ref{eq:overlap}:
\begin{align}
\langle  A  B \rangle &= \sum_i \int A_i B(\alpha) W_i(\alpha) \mathrm{d}^2 \alpha
\end{align}
The overlap integral used in this calculation is similar to descriptions of the standard Wigner function \cite{Cahill:1969jg,book:haroche06}. Shown in Fig.~\ref{fig:pauli_set} is a comparison between observables calculated by Eq.~\ref{eq:overlap} and those determined from a density matrix reconstruction.

\bs{Detector efficiency:} Under experimental conditions, the measured joint Wigner function is determined with point-by-point measurements of the joint observable $\langle \sigma_i P_\alpha \rangle$. Detector inefficiency results in a reduced visibility $\mathcal{V} \in [0,1]$ and in turn a reduced contrast of the measured joint Wigner functions $W^\mathrm{meas}_i(\alpha) = \mathcal{V} W^\mathrm{ideal}_i(\alpha)$. We can determine $\mathcal{V}$ by tracing over both the qubit and cavity states and comparing this to its ideal value $\int  W^\mathrm{ideal}_I(\alpha) \mathrm{d}^2 \alpha$ = 1:
\begin{align}
\mathcal{V} = \int W^\mathrm{meas}_I(\alpha) \mathrm{d}^2 \alpha
\end{align}
where $I$ is the qubit state identity operator. We observe $\mathcal{V} = 85\%$ and attribute this primarily to readout infidelity and qubit decay between the sequential measurements (See Sec. \ref{sec:errors}).

\bs{Density matrix reconstruction:} In Fig.~2 of the main text, we show the reconstructed density matrix of a target Bell-cat state $\ket{\psi_\mathrm{B} }$. We perform this reconstruction with a-priori assumptions that the cavity state is truncated to twelve occupied photon number states $N_\textrm{max} = 12$, the resulting noise of each averaged measurement is Gaussian distributed, and the reconstructed density matrix is positive semidefinite with trace equal to one.

Under these constraints, we perform a least squares regression using a Maximum likelihood estimation \cite{Smolin:2012vp}. To analyze this regression, we perform residual boostrapping on the reconstructed data set giving bounds on the error statistics of the inferred state.


\section{Encoded subspace}

%
\bss{Orthogonality of logical states: }In the main text, we describe encoded qubit states of the cavity where logical states $\ket{0_\mathrm{L}}, \ket{1_\mathrm{L}}$ correspond to  coherent states $\ket{\beta},\ket{-\beta}$. This approximation only holds for coherent states $\ket{\pm \beta}$ that are quasi-orthogonal $|\langle -\beta | \beta \rangle |^2 \ll 1$. To be more precise, we can calculate the maximum Von-Neumann entropy of the encoded space to determine its capacity to store information:
\begin{align}
S &= -\mathrm{Tr}\left[ \rho_\mathrm{max} \log_2{\rho_\mathrm{max}} \right] \nonumber \\
  &= -\sum_i \eta_i \log_2{\eta_i}
\end{align}
where $\rho_\mathrm{max} = \tfrac{1}{2}(\ket{\beta}\bra{\beta} + \ket{-\beta}\bra{-\beta})$ is the density matrix for a complete mixture of the logical subspace and $\eta$ is its set of eigenvalues. Rewriting $\rho_\mathrm{max}$ in the even/odd cat state basis:
\begin{equation}
\rho_\mathrm{max} = \tfrac{1}{2}(1+e^{-2|\beta|^2})\ket{\mathrm{E}}\bra{\mathrm{E}} + \tfrac{1}{2}(1-e^{-2|\beta|^2})\ket{\mathrm{O}}\bra{\mathrm{O}}
\end{equation}
where $\ket{\mathrm{E}},\ket{\mathrm{O}} = \frac{1}{\sqrt{2(1\pm e^{-2|\beta|^2})}} (\ket{\beta} \pm \ket{-\beta})$. Recall that $\braket{E|O} = 0$ for all coherent state amplitudes $\beta$. This gives the following entropy relation:
\begin{align}
    \label{eq:entropy}
S = - \tfrac{1}{2}(1+&e^{-2|\beta|^2}) \log_2\left( \tfrac{1}{2}(1+e^{-2|\beta|^2}) \right) \\ \nonumber
&- \tfrac{1}{2}(1-e^{-2|\beta|^2}) \log_2\left( \tfrac{1}{2}(1-e^{-2|\beta|^2}) \right)
\end{align}
Shown in Fig.~\ref{fig:entropy} is the capacity to store information using this encoding scheme. Entropy varies from zero bits to a value asymptotically approaching a single bit with increasing coherent state amplitudes $\beta$. The orthogonality between logical states $|\langle \beta | -\beta \rangle|^2$ is directly related to this information capacity and serves as a proxy for validating the qubit approximation of the produced cavity state.

\subsection{Encoded state observables}
The coherent state basis chosen in this report to represent the encoded qubit has Pauli operators:
\begin{align}
X_c &= \ket{{-\beta}}\bra{\beta} + \ket{\beta}\bra{{-\beta}} \\ \nonumber
Y_c &= j\ket{{-\beta}}\bra{\beta} - j\ket{\beta}\bra{{-\beta}} \\ \nonumber
Z_c &= \ket{\beta}\bra{\beta} - \ket{{-\beta}}\bra{{-\beta}} \\ \nonumber
I_c &= \ket{\beta}\bra{\beta} + \ket{{-\beta}}\bra{{-\beta}}
\end{align}
Here, we will show that the operators expressed in Eq.~4 of the main text, approximate these encoded Pauli operators. Assuming $\braket{\beta|{-\beta}} \ll 1$, we have the following photon-number parity $P$ relations:
\begin{align}
\bra{\beta}P_0\ket{\beta} &= \braket{\beta|{-\beta}} \ll 1 \\ \nonumber
\bra{\beta}P_0\ket{{-\beta}} &= \braket{\beta|{\beta}} = 1 \\ \nonumber
\bra{\beta}P_{\alpha}\ket{{\beta}} &= \braket{\beta-\alpha|{\alpha - \beta}} \ll 1 \\ \nonumber
\bra{\beta}P_{\alpha}\ket{{-\beta}} &= e^{2(\alpha\beta^* - \alpha^*\beta)}\braket{\alpha|{-\alpha}} \\ \nonumber
\end{align}
where $P_\alpha = D_\alpha P D_{-\alpha}$ for some displacement amplitude $\alpha$. Now taking the projector $M = \ket{\beta}{\bra{\beta}} + \ket{{-\beta}}\bra{{-\beta}}$, we derive the encoded state's Pauli Operators from the cavity state observables reported in Eq.~4 of the main text:
\begin{align}
M P_0 M^\dagger \approx \ket{{-\beta}}\bra{\beta} + \ket{\beta}\bra{{-\beta}} \\ \nonumber
M P_\beta M^\dagger \approx \ket{\beta}\bra{\beta} \\ \nonumber
M P_{-\beta} M^\dagger \approx \ket{{-\beta}}\bra{{-\beta}} \\ \nonumber
M P_{\frac{j\pi}{8\beta}} M^\dagger \approx j\ket{{-\beta}}\bra{\beta} - j\ket{\beta}\bra{{-\beta}}
\end{align}
Putting these relationships together, as in Eq.~4, builds the encoded state observables $\{X_c, Y_c, Z_c, I_c\}$ and reveals that these observables can be efficiently measured using Wigner tomography. $I_c$ and $Z_c$ require a comparison between two different observables. For true single-shot readout of these logical observable $Z_c$, measuring a single value in the cavity state Husimi-Q distribution $Q(\beta) = \frac{1}{\pi}\braket{\beta|\rho|\beta}$ can be employed where $Z_c = 2\pi Q(\beta) - 1$. This is being explored in future experiments.

\begin{figure*}[h]
\centering
\includegraphics[]{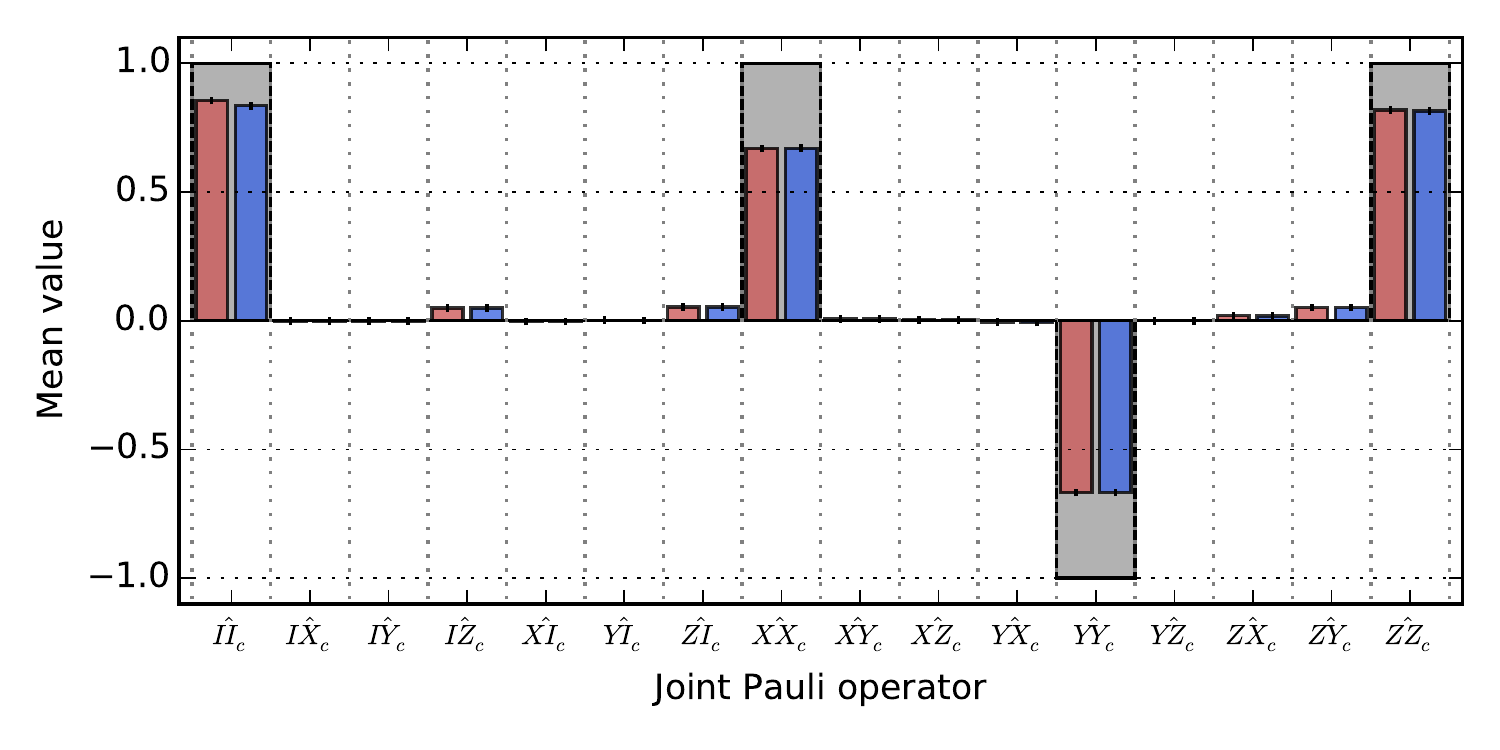}
\caption{\textbf{Reconstructed Pauli sets.}
The set of sixteen joint Pauli operators span the two-qubit Hilbert space of the qubit/encoded-qubit state. Shown is the Pauli set for the entangled target state $\ket{\psi_B}$ derived in two ways. (Red) is the reconstructed Pauli set using a density matrix reconstruction of the full quantum state with no normalization constraint, then projecting onto the encoded subspace. (Blue) shows the values discerned from an overlap integral of the measured joint-Wigner functions (Eq.~\ref{eq:overlap}). These measurements agree with each other within statistical errors.
}
   \label{fig:pauli_set}
\end{figure*}

\begin{figure*}[h]
\begin{floatrow}
\ffigbox{
\centering
\includegraphics[]{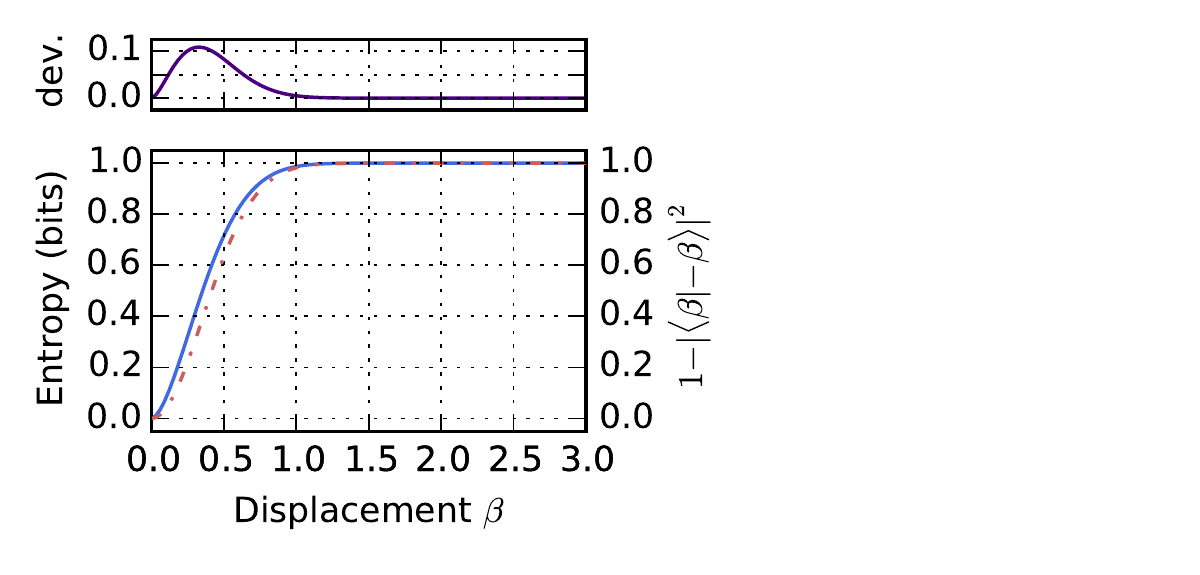}
}{
\caption{\textbf{Informational entropy.}
The capacity to store information into a cat state is determined by the orthogonality of its logical states $\ket{\beta}, \ket{-\beta}$. Shown is a comparison between the coherent state overlap (dashed line) and the maximum Von Neumann entropy Eq.~\ref{eq:entropy} (solid line) for this logical encoding. Notice that entropy rapidly approaches one bit for $\beta>1$, ensuring that information can be reliably encoded into the coherent states with manageable separations.}
   \label{fig:entropy}
}
\ffigbox{
\centering
\includegraphics[]{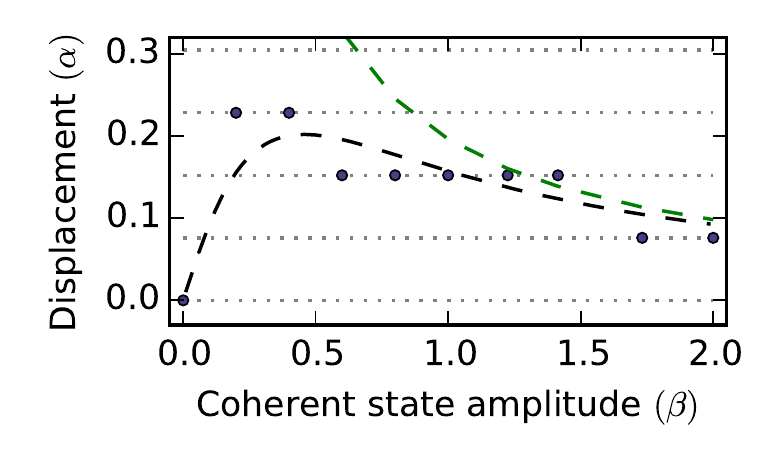}
}{
\caption{\textbf{Optimal displacement for Bell violation.}
For performing Bell test \#2, the optimal observables to measure maximum correlations depend on the size of the Bell-cat state Eq.~\ref{eq:optimal}. The dashed black line shows numerically calculated optimal displacement points as a function of coherent state amplitude $\beta$. Shown in circles are the experimentally determined optimal displacement values used to measure a maximum Bell violation. Differences between chosen and ideal values are a result of the discretization of our measurement settings. The dashed green line is the approximate trend $|\alpha_\mathrm{opt}| = |\frac{j\pi}{16\beta}|$ for large cat states, which diverge at small $\beta$.}
    \label{fig:optimal}
}
\end{floatrow}
\end{figure*}

\bs{Encoded state Pauli set:} We can represent the two qubit Bell state shown in Fig.~2 as a list of two-qubit correlations. The complete set constitutes the permutation of each of the single qubit Pauli set $\{  I,  X,  Y,  Z \}$. We can determine the two-qubit Pauli set from the complete reconstructed qubit-cavity state and projecting onto the encoded basis of $\{  I_c,  X_c,  Y_c,  Z_c \}$. Fig.~\ref{fig:pauli_set} shows the resulting two-qubit Pauli set for the transmon qubit and an encoded qubit  in the cavity mode, a variant of the reduced density matrix representation shown in Fig.~2 of the main text.

\bs{Encoded state preparation: }We can diagnose errors that can occur during state preparation from the reconstructed Pauli set. The dominant nonideal effects we explore are qubit decay during preparation and single-qubit rotation error.

During state preparation, the product state $\ket{\psi} = \frac{1}{\sqrt{2}}(\ket{g} + \ket{e})\otimes \ket{\beta}$ is initialized. Under the dispersive interaction, the system evolves into the entangled state $\frac{1}{\sqrt{2}}(\ket{g,\beta} + \ket{e, -\beta})$. To describe the effects of $T_1$ decay, we can look at the diagonal elements of the reduced density matrix after the entangling evolution of the dispersive interaction:
\begin{align}
\mathrm{diag}[\rho] = \tfrac{1}{2} \big\{ \ket{g,\beta}&\bra{g,\beta} + e^{-\gamma} \ket{e,-\beta}\bra{e,-\beta} \\ \nonumber
&+ \sum_k C_k \ket{g,\alpha_k}\bra{g,\alpha_k} \big\}
\end{align}
where $\alpha_i = \beta e^{j\chi t_k}$ represents coherent states when a jump occurred at time $t_k$ and $\gamma = \frac{\pi}{\chi T_1}$. Projecting onto the logical basis (here we will approximate $|\langle \alpha_k | \beta \rangle |^2 \ll 1$) produces the resulting scaling on the joint Pauli measurements for the ideal state:
\begin{equation}
\begin{array}{rr}
\braket{II_c} \propto   \frac{1}{2} \left(1 + e^{-\gamma} \right) &
\braket{ZI_c} \propto   \frac{1}{2} \left(1 - e^{-\gamma} \right) \\[1em]
\braket{ZZ_c} \propto   \frac{1}{2} \left(1 + e^{-\gamma} \right) &
\braket{IZ_c} \propto   \frac{1}{2} \left(1 - e^{-\gamma} \right)
\end{array}
\end{equation}
This gives us an approximate method to predict our ability to prepare a state that is within the logical subspace given our experimental parameters, $\braket{II_c} = 0.99$. Our measured value taking into account detector inefficiencies produces $\braket{II_c} = 0.98$.

The preparation of this entangled system is also sensitive to the amplitude of the initial qubit rotation $Y/2 = R^{\frac{\pi}{2}}_{\hat y}$. The angle of rotation $\theta$ will determine the prepared state as:
\begin{equation}
\ket{\psi} = \mathcal{N} \{ \cos{\tfrac{\theta}{2}} \ket{g,\beta} + \sin{\tfrac{\theta}{2}} \ket{e,-\beta} \}
\end{equation}
For the states prepared nearly as $\ket{\psi_\mathrm{B}}$, $\theta \approx \frac{\pi}{2}(1 - \delta \theta)$ and will result in the following modification of the joint Pauli measurements for an ideal state:
\begin{equation}
\begin{array}{rr}
\braket{ZI_c} \propto  \frac{\pi}{4}\delta \theta \\[1em]
\braket{IZ_c} \propto   \frac{\pi}{4}\delta \theta
\end{array}
\end{equation}
From the measurements in Fig.~\ref{fig:pauli_set}, we can determine that the relative error for the rotation angle in our preparation rotation to be $\delta \theta \approx 2.8 \%$.


\section{Bell test analysis}

%
\label{sec:bell}
The main text reports CHSH Bell tests composed of two qubit observables $A,B$ and two cavity observables $A_c,B_c$, correlated such that:
\begin{equation}
\mathcal{O} = \mean{AA_c} + \mean{AB_c} - \mean{BA_c} + \mean{BB_c}
\end{equation}
We perform two variants of this test on the state $\ket{\psi_\mathrm{B}}$.

\bs{Test \#1 Model: }In the first test we choose qubit cavity observables $Z_c, X_c$ and qubit observables $Z(\theta), X(\theta)$ where:
\begin{align}
Z(\theta) = Z\cos{\tfrac{\theta}{2}} - X\sin{\tfrac{\theta}{2}} &&
X(\theta) = X\cos{\tfrac{\theta}{2}} + Z\sin{\tfrac{\theta}{2}}
\end{align}
This angle $\theta$ corresponds to a rotation of the qubit state before detection. In Fig.~4a, we plot $\mathcal{O}$ for each of the four permutations of the joint observables and find a maximum Bell violation for an angle $\theta = -\tfrac{\pi}{4}$ giving observables:
\begin{equation}
    \begin{array}{ll}
    A = \frac{X + Z}{\sqrt{2}};  &  B = \frac{X - Z}{\sqrt{2}} \\
    A_c = Z_c;                   &  B_c = X_c
    \end{array}
\end{equation}
As shown in Fig.~4c, we can model the effects of photon loss and measurement inefficiency on the maximum violation. For the ideal case, an overlap of the coherent state superposition reduces  contrast in $\mean{AZ_c}$ and $\mean{BZ_c}$ and will limit the maximum Bell signal:
$$\mathcal{O}_\mathrm{ideal} = \sqrt{2}( 2 - e^{-8|\beta|^2} )$$
Measurement inefficiency will reduce the contrast of this maximum Bell signal which we expect to go as the visibility $\mathcal{V}$:
$$\mathcal{O}_\mathrm{vis} = \sqrt{2}\mathcal{V}( 2 - e^{-8|\beta|^2} )$$
Photon loss will also have an effect on the maximum Bell signal by reducing the measured contrast of all correlations for $\mean{AX_c}$ and $\mean{BX_c}$. This produces the an amplitude dependent maximum Bell Signal:
$$\mathcal{O}_\mathrm{loss} = \sqrt{2}( 1 - e^{-8|\beta|^2} - e^{-2|\beta|^2 \gamma})$$
where $\gamma = \frac{t_\mathrm{eff}}{\tau_s}$ such that $\tau_s$ is the photon decay time constant and $t_\mathrm{eff}$ is the effective time to create and measure the Bell-cat state. Finally taking into account both visibility and photon loss produces the expected maximum Bell signal:
$$\mathcal{O}_\mathrm{pred} = \sqrt{2}\mathcal{V}( 1 - e^{-8|\beta|^2} - e^{-2\gamma|\beta|^2}) $$
This predicted Bell signal is shown in Fig.~4 of the main text using the measured joint-Wigner contrast $\mathcal{V} = 0.85$ and time between cavity state creation and detection $t_\mathrm{eff} = 1.24 ~\mu \mathrm{s}$.

\bs{Test \#2 Model: }In the second test, we choose qubit observables $X, Y$ and cavity observables $X_c(\alpha), Y_c(\alpha)$ where:
\begin{align}
X_c(\alpha) = D_{j\alpha} P_0 D^\dagger_{j\alpha} \approx X_c\cos{\tfrac{\alpha}{4\beta}} + Y_c\sin{\tfrac{\alpha}{4\beta}} \\ \nonumber
Y_c(\alpha) = D_{j\alpha} P_{\frac{j\pi}{8\beta}} D^\dagger_{j\alpha} \approx Y_c\cos{\tfrac{\alpha}{4\beta}} - X_c\sin{\tfrac{\alpha}{4\beta}}
\end{align}
Where the displacement amplitude $\alpha$ corresponds to an approximate rotation of the encoded cavity state before detection. In Fig.~4b, we plot $\mathcal{O}$ for each of the four permutations of the joint observables and find a maximum Bell violation for a displacement $\alpha = 0.15$ for $\beta = 1$ which produces the approximate observables:
\begin{equation}
    \begin{array}{ll}
    A = X;                       &  B = Y\\
    A_c = \frac{X_c + Y_c}{\sqrt{2}}   &  B_c = \frac{X_c - Y_c}{\sqrt{2}}
    \end{array}
\end{equation}
Shown in Fig.~4c, we can also model the effects of photon loss and measurement inefficiency for the second test. The ideal case is the result of four summed joint Wigner values represented as:
$$ \mathcal{O}_\mathrm{ideal} = 2(\cos{4\alpha_0 \beta} + \sin{4\alpha_0 \beta})e^{-2|\alpha_0|^2} $$
where $\alpha_0$ is an optimal displacement for maximum violation which can be calculated from Eq.~\ref{eq:optimal} and in detail in Ref.~\cite{Park:2012iw}. Taking into account photon loss and measurement inefficiency produces the following relationship:
$$ \mathcal{O}_\mathrm{pred} = 2\mathcal{V}e^{-2\gamma|\beta|^2}(\cos{4\alpha_0 \beta} + \sin{4\alpha_0 \beta})e^{-2|\alpha_0|^2} $$
This predicted Bell signal is shown in Fig.~4b of the main text using the measured joint-Wigner contrast $\mathcal{V} = 0.85$ and an effective time $t_\mathrm{eff} = 1.24~\mu\mathrm{s}$.

\bs{Optimal measurements for encoded observables: } Eq.~3 of the main text describes the ideal observables to efficiently determine an encoded qubit state observable using a superposition state with $|\beta| \gg 1$. In fact, the optimal measurement for particular observables will be further modified for smaller coherent displacements.

For the second CHSH experiment, the optimal observable $P_{\pm j \alpha_0} \sim \frac{1}{\sqrt{2}}(\hat X_c \pm \hat Y_c)$ follows the relation:
\begin{equation}
    \label{eq:optimal}
\frac{\beta - \alpha_0}{\beta + \alpha_0} = \tan{4\alpha_0\beta}
\end{equation}
where $\alpha_0$ is the amplitude for a coherent displacement $D_{j\alpha_0}$ to perform the measurement $P_{j \alpha_0}$ given $\beta$. Further details are discussed in Ref.~\cite{Park:2012iw}. In the large $\beta$ limit, the observable corresponds to the encoded qubit state observable $\frac{1}{\sqrt{2}}(\hat X_c + \hat Y_c)$ and follows the relationship $P_{\alpha = \frac{j\pi}{16 \beta}}$ as related in Eq.~4 of the main text. Shown in Fig.~\ref{fig:optimal} is the predicted and chosen optimal values for a maximum CHSH Bell signal.

\begin{figure*}[h]
\centering
\includegraphics[]{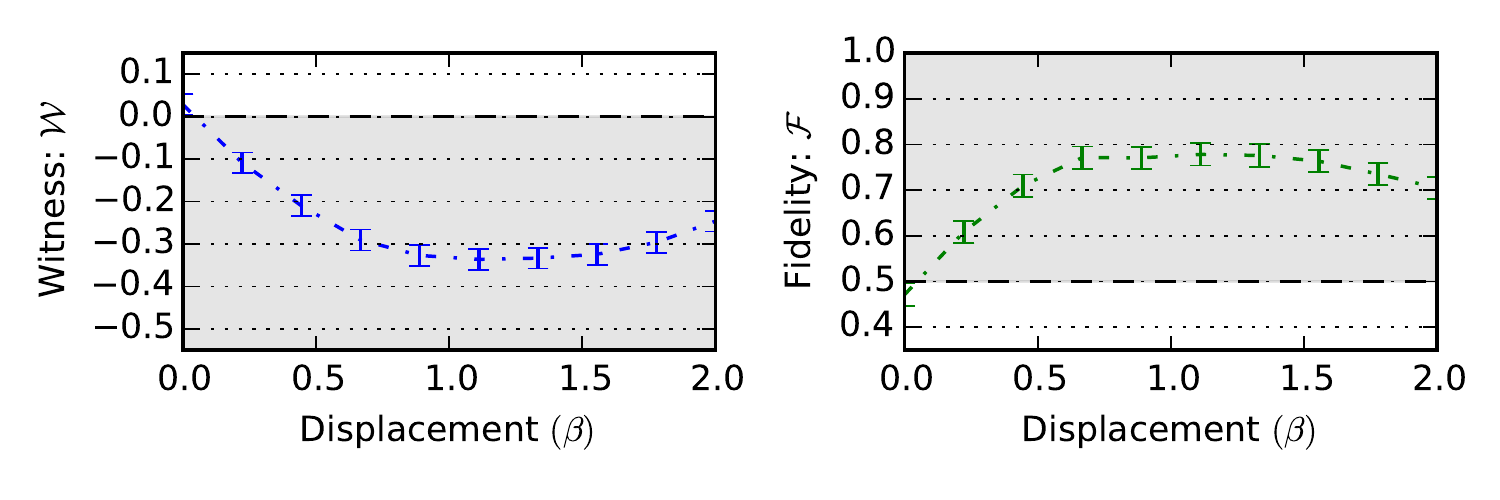}
\caption{\textbf{Entanglement witnesses with cat states.}
An entanglement witness and direct fidelity estimation (DFE) are determined by measuring four qubit-cavity correlations. (a) The entanglement witness $\mathcal{W} = II - ZZ - XX + YY$ shows entanglement for all negative values (grey shading). (b) DFE to a target Bell state $\mathcal{F} = II + XX - YY + ZZ$ is also shown where entanglement can be confirmed for values above $\mathcal{F} >0.5$. Notice that these two witnesses have a much looser bound for entanglement than the CHSH Bell test.
}
   \label{fig:witness}
\end{figure*}

\bs{Two-qubit entanglement witnesses:} Two qubit entanglement can also be quantified by an entanglement witness $\mathcal{W} = II_c -XX_c + YY_c - ZZ_c$~\cite{Horodecki:2009gb} for a Bell state $\ket{\psi} = \frac{1}{\sqrt{2}} (\ket{gg} + \ket{ee})$. The witness `confirms' entanglement for all observations of $\braket{\mathcal{W}} < 0$. Shown in Fig.~\ref{fig:witness}, we report $\mathcal{W}$ (as well as its corresponding direct fidelity estimation $\mathcal{F}$) as a function of coherent state amplitude $\beta$ using the optimal displacements described in Fig.~\ref{fig:optimal}. As expected, entanglement is not detected for a $\beta = 0$ coherent state (a product state $\frac{1}{\sqrt{2}} (\ket{g} + \ket{e}) \otimes \ket{0}$).

\begin{figure*}[h]
\begin{floatrow}
\ffigbox{
\label{fig:settings}
\centering
\includegraphics[]{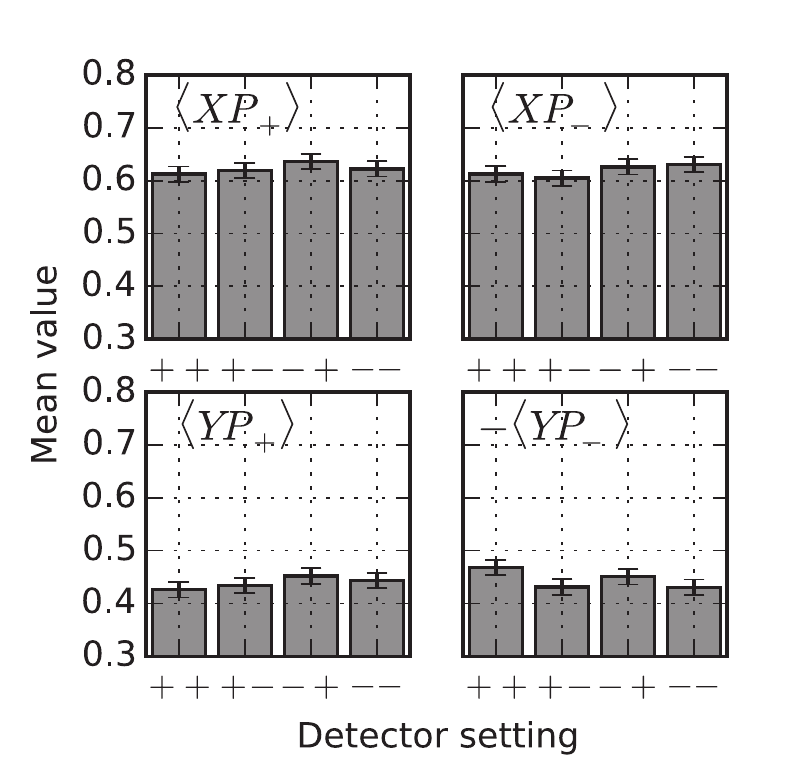}
}{
\caption{\textbf{Observables from each detector setting.}
To ensure that a particular detector setting is not producing systematic errors we have not taken into account. We report a Bell test for each detector setting used to observe our maximum violation in test \#2. The expectation value of each observable used in that Bell test is shown for the four detector settings used Sec.~\ref{sec:errors}. Significant deviations due to unexpected systematic errors are not observed.}
}
\ffigbox{
\label{fig:bell}
\centering
\includegraphics[]{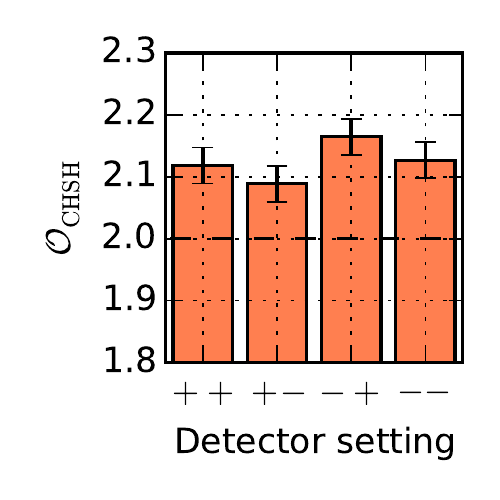}
}{
\caption{\textbf{Bell test for each detector setting.}
A Bell test is analyzed for each detector setting to determine the effects of possible systematic errors. Each of these subtests violate Bell's inequality by more than three standard deviations of their statistical error.}
}
\end{floatrow}
\end{figure*}

\bs{Bell test for each detector setting: }We analyze the systematic errors that can occur from a particular detector setting. Shown in Fig.~\ref{fig:settings} are the observables used to calculated a Bell violation using test \#2 for each of the four detector settings Sec.~\ref{sec:errors}. Systematic errors are shown to be within stastical bounds of the experiment and each detector setting violates Bell's inequality by at least three standard deviations, see Fig.~\ref{fig:bell}. In the main text, we report measurements from the combined data set resulting in smaller statistical error and a stronger violation of Bell's inequality.


\section{Quantum measurement back-action}

The sequential measurement protocol allows us to observe the result of quantum measurement back-action of the qubit on the cavity state. The result of an ideal qubit measurement outcome $M_m$ will give a projected qubit-cavity state:
\begin{equation}
\ket{\psi_\mathrm{m}} = \frac{ M_m\ket{\psi} }
        { \sqrt{ \langle \psi | M^\dagger_m M_m | \psi \rangle} }
\end{equation}
Measuring along the $\{ X,Y,Z \}$ axes of the qubit gives three measurement sets:
\begin{align}
    \begin{array}{rc|c}
    X: &  \frac{1}{2}
      \begin{pmatrix}
        1  &  1\\
        1  &  1
      \end{pmatrix}\otimes\mathbbm{1}_c, &
    \frac{1}{2}
      \begin{pmatrix}
        1  &  -1\\
        -1  &  1
      \end{pmatrix}\otimes\mathbbm{1}_c \\[1.5em]
    Y: &  \frac{1}{2}
      \begin{pmatrix}
        1  &  -j\\
        j  &  1
      \end{pmatrix}\otimes\mathbbm{1}_c, &
    \frac{1}{2}
      \begin{pmatrix}
        1  &  j\\
        -j  &  1
      \end{pmatrix}\otimes\mathbbm{1}_c \\[1.5em]
    Z: &
      \begin{pmatrix}
        1  &  0\\
        0  &  0
      \end{pmatrix}\otimes\mathbbm{1}_c, &
      \begin{pmatrix}
        0  &  0\\
        0  &  1
      \end{pmatrix}\otimes\mathbbm{1}_c \\
    \end{array}
\end{align}

\begin{figure*}[h]
\begin{center}
\includegraphics[scale=1.0]{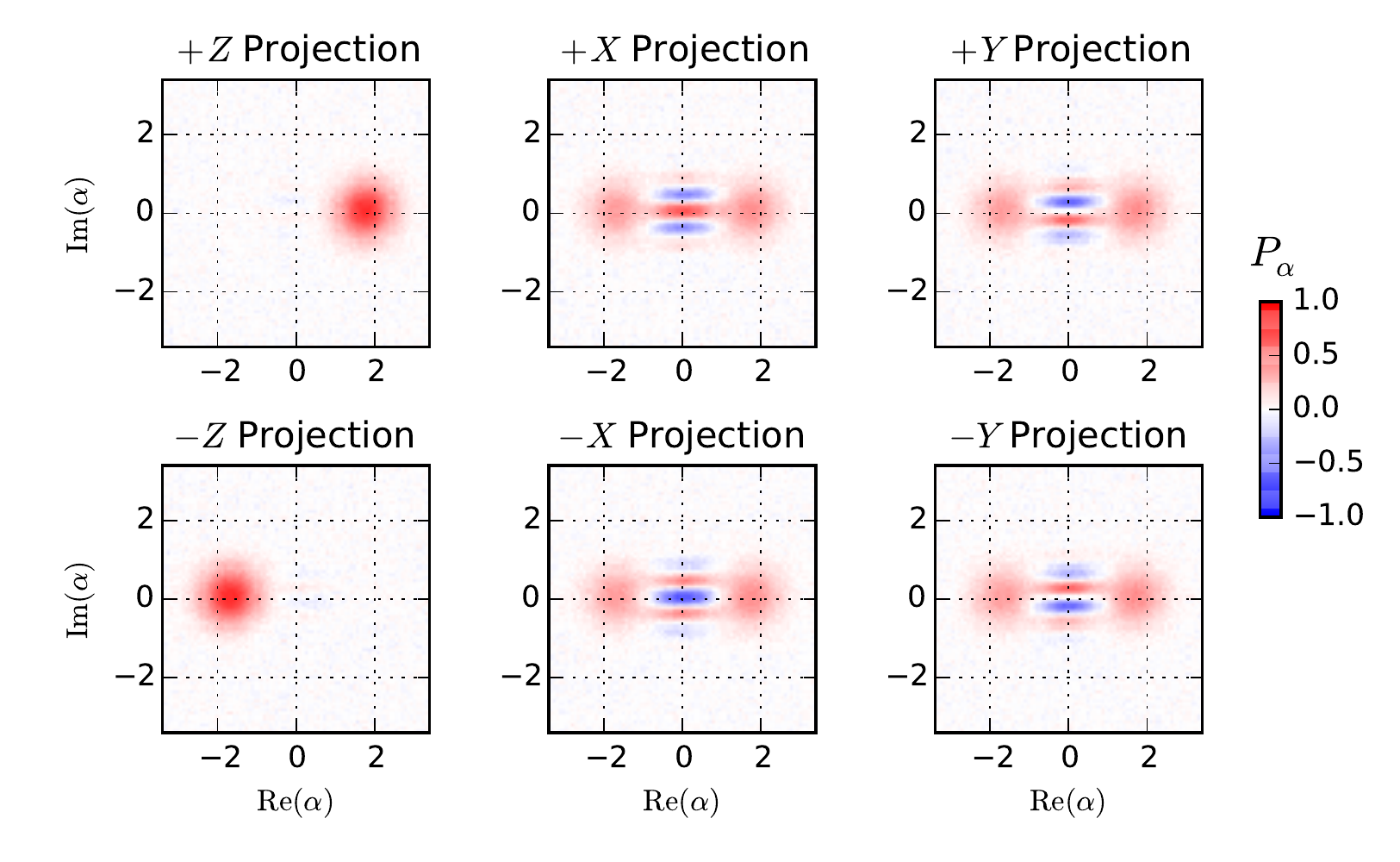}
\caption{\textbf{Qubit measurement back-action of a Bell-cat state.} The resulting projections of the state $\ket{\psi_\mathrm{B}} = \frac{1}{\sqrt{2}} (\ket{g, \beta} + \ket{e, -\beta})$ due to a particular qubit measurement outcome. Note that measuring along the $X$ and $Y$ axes results in a projected cat state each with different superposition phases. Combining these measurements with the probability to obtain each result describes the entire system and is used to create the joint Wigner function representation in Fig.~2 of the main text.
}
   \label{fig:projections}
   \end{center}
\end{figure*}

\bs{Bell-cat projections:} We prepare the system in a Bell-cat state as in Eq.2, and measure along each of the three qubit axes. These three measurements results in six possible outcomes $\ket{\psi_\mathrm{m}} = \ket{\psi_q}\otimes \ket{\psi_c}$ with the projected cavity states:
\begin{align}
\begin{array}{rrc|c}
    \ket{\psi_\mathrm{cav}} \to & X: &  \mathcal{N}\left( \ket{\beta} + \ket{\beta} \right)
                                &       \mathcal{N}\left( \ket{\beta} - \ket{\beta} \right)   \\
                                & Y: &  \mathcal{N}\left( \ket{\beta} - j\ket{\beta} \right)
                                &       \mathcal{N}\left( \ket{\beta} + j\ket{\beta} \right)  \\
                                & Z: &  \ket{\beta}
                                     &  \ket{-\beta}
\end{array}
\end{align}
See Fig.~\ref{fig:projections} for each projective measurement of the Bell-cat state $\ket{\psi_\mathrm{B}}$. The method of using strong projective measurements to create cat states has been demonstrated in previous experiments~\cite{Deleglise:2008gt}.

\begin{figure*}[h]
\begin{center}
\includegraphics[scale=1.0]{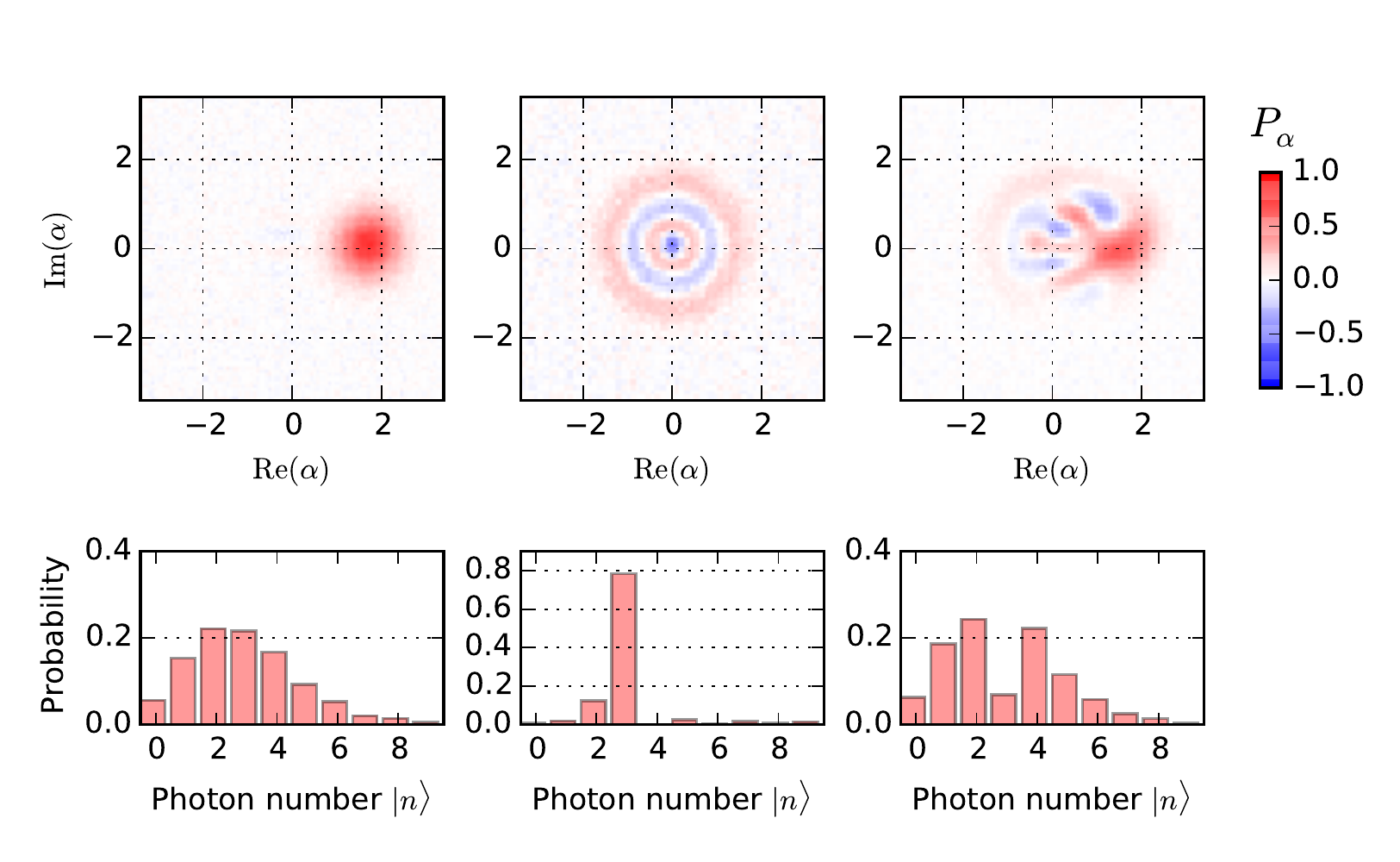}
\caption{\textbf{Qubit measurement back-action for an entangled Fock state.}
(a) A measured Wigner function of a coherent state $\ket{\beta}$ where $\beta = \sqrt{3}$ results in a Poissonian photon distribution. Performing a photon-selective qubit rotation on the $m^\mathrm{th}$ level where $m = 3$ results in an entangled state $\ket{\psi} = C_m\ket{e, m} + \sum_{n \neq m} C_n \ket{g, n}$ where $C_n$ is the coefficient of the $n^\mathrm{th}$ photon number state $C_n = \langle n | \beta \rangle$. (b) The measured Wigner function of the cavity state after the qubit has been measured in the $-Z$ state results in a 3-photon Fock state. (c) Instead, when a $+Z$ result is obtained the measured cavity state Wigner function is a Fock-state subtracted coherent state $\ket{\psi_c} = \mathcal{N} \sum_{n \neq 3} C_n \ket{n}$.}
   \label{fig:fock_proj}
   \end{center}
\end{figure*}

\bs{Fock state projections:} We prepare the system in a state such that the qubit state $\ket{e}$ is correlated with the $m^\mathrm{th}$ photon Fock state $\ket{m}$ of a coherent state $\ket{\beta}$ (in this example $m=3$ photons and $\beta = \sqrt{3}$). This can be written as:
\begin{align}
\ket{\psi} = C_m\ket{e, m} + \sum_{n \neq m} C_n \ket{g, n}
\end{align}
where $C_m = \langle m | \beta \rangle$. Shown in Fig.~\ref{fig:fock_proj}, when the qubit is measured along the $\hat Z$ axis we observe a change in photon statistics such that a $+1$ event projects the cavity onto the state $\ket{\psi_\mathrm{cav}} =  \mathcal{N}\left( \ket{\beta} - C_m \ket{m} \right)$ and a $-1$ event projects onto the Fock state $\ket{\psi_\mathrm{cav}} = \ket{m}$.

\end{document}